\documentclass[usenatbib,useAMS]{mn2e}
\bibpunct{(}{)}{;}{a}{}{,}

\usepackage{enumerate}

\usepackage{aas_macros}

\usepackage{amsmath}
\usepackage{mathtools}
\usepackage{IEEEtrantools}
\usepackage{float}

\usepackage{txfonts}

\usepackage[applemac]{inputenc}
\usepackage[T1]{fontenc} 

\usepackage[english]{babel}
\usepackage{booktabs}
\usepackage{multirow}

\usepackage{savesym}
\savesymbol{iint}
\savesymbol{iiint}
\savesymbol{iiiint}
\savesymbol{idotsint}

\usepackage{amssymb,amsmath}

\usepackage{amstext}
\usepackage{subfigure}
\usepackage{graphicx}
\usepackage[np,noautolanguage]{numprint}

\usepackage{bigstrut}

\usepackage{url}
\usepackage{hyperref}
\usepackage{pdfsync}
\bibpunct{(}{)}{;}{a}{}{,} % to follow the A&A style

\usepackage{threeparttable}

\newcommand{\colhspace}{0.02\hsize}

\def\lb{\left[}
\def\rb{\right]}

\newcommand{\ave}[1]{\langle #1 \rangle}

\newcommand{\txn}[1]{\textnormal{#1}}

\newcommand{\gadget}{\textsc{gadget2}}
\newcommand{\adaptahop}{\textsc{adaptahop}}
\newcommand{\treemaker}{\textsc{treemaker}}

\newcommand{\Planck}{\textit{Planck}}

\newcommand{\LCDM}{\hbox{$\Lambda\txn{CDM}$}}

\newcommand{\MpcInv}{\hbox{$\txn{Mpc}^{-1}$}}

\newcommand{\hInv}{\hbox{$h^{-1}$}}
\newcommand{\hInvMpc}{\hbox{$\hInv\txn{Mpc}$}}

\newcommand{\M}{\txn{M}}

\newcommand{\Lya}{\hbox{Ly\,$\alpha$}}

\newcommand{\Mstar}{\hbox{$\txn{M}_{\ast}$}}
\newcommand{\Msun}{\hbox{$\txn{M}_{\sun}$}}

\newcommand{\Fgrowth}{\hbox{$\txn{F}_\txn{growth}$}}
\newcommand{\Fphys}{\hbox{$\txn{F}_\txn{phys}$}}

\newcommand{\Mhalo}{\hbox{$\txn{M}_\txn{halo}$}}
\newcommand{\Mvir}{\hbox{{$\txn{M}_\txn{halo}$}}}
\newcommand{\MvirP}{\hbox{$\txn{M}_\txn{vir}^\txn{peak}$}}

\newcommand{\Tvir}{\hbox{{$T_\txn{vir}$}}}

\newcommand{\Muv}{\hbox{$M_\txn{FUV}$}}

\newcommand{\Auv}{\hbox{$A_\txn{FUV}$}}

\newcommand{\MuvLim}{\hbox{$M_\txn{FUV}^\txn{lim}$}}

\newcommand{\Myr}{\hbox{$\txn{Myr}$}}

\newcommand{\hInvMsun}{\hbox{$\hInv \Msun$}}

\newcommand{\fNL}{\hbox{$f_\txn{NL}$}}

\newcommand{\fNLn}{\hbox{$f_\txn{NL}^0$}}

\newcommand{\FeH}{\hbox{$[\txn{Fe}/\txn{H}]$}}

\newcommand{\ndotion}{\hbox{$\dot{n}_{\rm ion}$}}
\newcommand{\QHII}{\hbox{$Q_{\rm HII}$}}
\newcommand{\nH}{\hbox{$\ave{n_{\mathrm{H}}}$}}
\newcommand{\nE}{\hbox{$n_{\mathrm{e}}$}}
\newcommand{\fE}{\hbox{$f_{\mathrm{e}}$}}
\newcommand{\mH}{\hbox{$m_{\mathrm{H}}$}}
\newcommand{\Omegam}{\hbox{$\Omega_{\mathrm{m}}$}}
\newcommand{\Omegab}{\hbox{$\Omega_{\mathrm{b}}$}}

\newcommand{\Omegal}{\hbox{$\Omega_{\Lambda}$}}
\newcommand{\rhoc}{\hbox{$\rho_{\mathrm{c}}$}}
\newcommand{\Xp}{\hbox{$X_{\mathrm{p}}$}}
\newcommand{\Yp}{\hbox{$Y_{\mathrm{p}}$}}
\newcommand{\alphaB}{\hbox{$\alpha_{\mathrm{B}}$}}

\newcommand{\CHII}{\hbox{$C_{\mathrm{HII}}$}}
\newcommand{\trec}{\hbox{$t_{\mathrm{rec}}$}}
\newcommand{\fesc}{\hbox{$f_{\mathrm{esc}}$}}
\newcommand{\xiion}{\hbox{$\xi_{\mathrm{ion}}$}}
\newcommand{\rhoUV}{\hbox{$\rho_{\mathrm{UV}}$}}
\newcommand{\tauE}{\hbox{$\tau_e$}}
\newcommand{\fb}{\hbox{$f_{\mathrm{b}}$}}

\title[Effect of primordial non-Gaussianities on the far-UV LF of high-redshift galaxies]{Effect of primordial non-Gaussianities on the far-UV luminosity function of high-redshift galaxies: implications for cosmic reionization}
\author[J. Chevallard]
{Jacopo~Chevallard$^{1}$\thanks{e-mail:chevalla@iap.fr},
Joseph Silk$^{1,2}$,
Takahiro~Nishimichi$^1$,
Melanie Habouzit$^1$,
\newauthor
Gary A. Mamon$^1$,
S\'ebastien~Peirani$^1$
\\
\\
$^{1}$UPMC-CNRS, UMR7095, Institut d'Astrophysique de Paris, F-75014, Paris, France\\
$^{2}$ Department of Physics and Astronomy, The Johns Hopkins University, Baltimore MD 20218, USA
}

\begin{document}

\date{Submitted to MNRAS on }

\maketitle

\label{firstpage}

% Abstract
\begin{abstract}

Understanding how the intergalactic medium (IGM) was reionized at $z\gtrsim6$ is one of the big challenges of current high redshift astronomy. It requires modelling the collapse of the first astrophysical objects (Pop III stars, first galaxies) and their interaction with the IGM, while at the same time pushing current observational facilities to their limits. The observational and theoretical progress of the last few years have led to the emergence of a coherent picture in which the budget of hydrogen-ionizing photons is dominated by low-mass star-forming galaxies, with little contribution from Pop III stars and quasars. The reionization history of the Universe therefore critically depends on the number density of low-mass galaxies at high redshift.
In this work, we explore how changes in the cosmological model, and in particular in the statistical properties of initial density fluctuations, affect the formation of early galaxies. Following \citet{Habouzit2014}, we run 5 different N-body simulations with Gaussian and (scale-dependent) non-Gaussian initial conditions, all consistent with \Planck\ constraints. By appealing to a phenomenological galaxy formation model and to a population synthesis code, we compute the far-UV galaxy luminosity function down to $\Muv = -14$ at redshift $ 7 \le z \le 15$. We find that models with strong primordial non-Gaussianities on $\lesssim$ Mpc scales show a far-UV luminosity function significantly enhanced (up to a factor of 3 at $z=14$) in low-mass galaxies.
We adopt a reionization model calibrated from state-of-the-art hydrodynamical simulations and show that such scale-dependent non-Gaussianities leave a clear imprint on the Universe reionization history and electron Thomson scattering optical depth \tauE. Although current uncertainties in the physics of reionization and on the determination of \tauE\ still dominate the signatures of non-Gaussianities, our results suggest that \tauE\ could ultimately be used to constrain the statistical properties of initial density fluctuations.

\end{abstract}

%Keywords

%TC:ignore

% Inizio testo

\section{Introduction}

Most baryons in the Universe exist in the form of ionized gas (mainly hydrogen and helium) in the intergalactic medium (IGM) \citep[e.g.][]{Fukugita1998,Cen2006}. The IGM plays the role of a gas reservoir by feeding dark-matter haloes with fresh gas available for star formation, in this way connecting the properties of large-scale structures to those of single haloes and galaxies. At very early times ($z\gtrsim20$), gas in the IGM is  neutral, and it remains neutral until the appearance of the first sources of ionizing radiation, namely metal-free Pop III stars, early galaxies and quasars. These sources can potentially provide enough photons to almost fully ionize the IGM by redshift $\sim6$ \citep[e.g.][]{Fan2006}. 

Although the details of Universe `reionization' are still uncertain, the last decade has seen an impressive improvement of our knowledge of this phase thanks to large observational and modelling efforts. On the one hand, very deep multi-wavelength images have allowed the detection of $\gtrsim 1500$ galaxy candidates at (photometric) redshift $z > 6$ \citep[e.g.][]{Oesch2013,Bouwens2014}. This has allowed the first statistical studies of galaxy populations at high redshift, e.g. the accurate determination of the ultraviolet galaxy luminosity function up to $z = 8$ \citep{Bouwens2014}. On the other hand, state-of-the-art hydrodynamic simulations are exploring, with improved resolution and refined physical recipes, the formation of the first galaxies and their contribution to the reionization process \citep[e.g.][]{So2014,Wise2014}.
At present, there is broad agreement on Universe reionization being driven by UV radiation emitted by hot, massive stars born in the first galaxies. The low number density of quasars at high redshift ($z\gtrsim6$) make their contribution to reionization negligible \citep[e.g.][]{Hopkins2007,Faucher2009}, while many computations have shown that the contribution from metal-free, population III stars is also of secondary importance \citep[e.g.][]{Paardekooper2013,Wise2014}. 
This means that a crucial ingredient for any reionization model is the number of ionizing photons emitted by early galaxies. This number depends, among other factors, on the number density of galaxies emitting UV radiation at different redshifts, that is on the evolution with time of the far-UV galaxy luminosity function. Several works so far have explored the effect on cosmic reionization of assumptions about the far-UV luminosity function of galaxies, such as the minimum mass of a halo able to sustain star formation and the shape of the faint-end of the luminosity function, assuming a fixed cosmological model. 

In this work, we adopt a different approach, exploring the effect of varying the cosmological model, and in particular the statistical properties of initial density perturbations. These perturbations, usually described by a Gaussian random field, evolve with time causing the collapse of dark matter particles, and at later times of baryons, into haloes. The assumption of Gaussian initial density perturbations is supported, at large scales, by the measurements of the temperature fluctuation of the cosmic microwave background (CMB) radiation. In particular, the recent analysis by the \Planck\ satellite puts a stringent constraint on the magnitude of a possible departure from Gaussianity using multipoles $\ell < 2500$, which corresponds to $k \lesssim 0.18 \, \MpcInv$ in comoving wave number \citep{Planck2013_24}. 

There is still, however, room for significant non-Gaussianities on smaller scales beyond the reach of CMB observations. Some inflationary models indeed predict the presence of scale-dependent non-Gaussian initial density perturbations (e.g., Dirac-Born-Infeld inflation: \citealt{Silverstein2004,Alishahiha2004,Chen2005}). It is possible for these models to pass the CMB constraints, while leaving a significant impact on  structure formation relevant to fluctuations on smaller scales. Although previous theoretical work has shown that the scale-dependence of non-Gaussianities can alter the abundance and clustering of collapsed objects, these studies mainly focus on the high-mass end of the mass function and at late times \citep{LoVerde2008,Shandera2011,Becker2011}, with the exception of the work by \citet{Crociani2009}, which focused on the reionization epoch. Given the already tight constraints on primordial non-Gaussianities on large scales, their imprint on smaller objects that formed at an earlier stage would be a complementary probe for understanding the statistical properties of the initial cosmic perturbations over a wider dynamic range.

\citet[][hereafter paper I]{Habouzit2014} have recently considered the effects of scale-dependent non-Gaussianity on the halo and stellar mass functions of galaxies. For this, four cosmological N-body simulations were run, each with different spectra for the scale-dependence of the local non-Gaussianity, plus an additional one with Gaussian initial conditions, all with the same phases. The level of non-Gaussianity was high on Galactic scales, but low enough to be consistent with the CMB constraints from the \Planck\ mission on cluster scales and larger. A simple galaxy formation model was applied to the halo merger trees run forward in time: this was based on the redshift-dependent stellar to halo mass relation of \citet{Behroozi2013}, which was modified to prevent stellar masses from decreasing in time. Paper~I concludes that the stellar mass function is significantly altered if the non-Gaussianity varies very strongly with scale.

In the present work, we analyse the same  cosmological simulations as in paper~I, extending the analysis of paper~I in several ways. We appeal to the simple, physically motivated, galaxy formation model of \citet{Mutch2013} to compute the stellar mass assembly in each dark matter halo. Next, we adopt the \citet{BC03} stellar population synthesis code to compute the far-UV luminosity function of galaxies in the redshift range $7\le~z~\le15$.  
Finally, we consider a `standard' reionization model and compute the reionization history of the Universe, exploring the impact of the different scale-dependent non-Gaussianities on reionization. 

In Section~\ref{sec:model} below, we describe the cosmological simulations along with the algorithms used to generate the non-Gaussian initial conditions, to identify haloes and build merger trees. We also present the simple galaxy formation model of \citet{Mutch2013}, as well as our additions to and modifications of the model. 
In Section~\ref{sec:results}, we show the effect of  different scale-dependent non-Gaussianities on the halo and stellar mass functions, and on the far-UV luminosity function of galaxies at redshift $7\le~z~\le15$. 
In Section~\ref{sec:reioniz_model}, we present the reionization model that we later adopt to compute the reionization history of the Universe.
In Section~\ref{sec:reioniz}, we quantify the impact of different levels of non-Gaussian initial density perturbations on the Universe reionization history and on the optical depth of electrons to Thomson scattering.
In Section~\ref{sec:uncert}, we discuss the assumptions of our reionization model and the possible sources of uncertainty.
Finally, we discuss in Section~\ref{sec:conclus} the implications of our findings on Universe reionization, and we highlight how more accurate measurements of the electron Thomson scattering optical depth can be used to constrain the level of primordial non-Gaussianities.

\section{Model}\label{sec:model}

We start by presenting the simulation setup, i.e. the algorithm we used to generate the (Gaussian and non-Gaussian) initial density perturbations and the N-body code adopted to compute their time evolution. We also introduce the codes we employ to identify the dark matter haloes at each time-step of the simulations and to build the merger trees.
We end this section by introducing the analytic galaxy formation model of \citet{Mutch2013}, along with our modification on the efficiency of baryon conversion into stars and addition of a prescription for the chemical evolution of galaxies, and the \citet{BC03} population synthesis code.
 
\subsection{Non-Gaussian initial density perturbations} 
 
The statistical properties of the initial density perturbations depend on the high energy physics adopted at the very first instants of the Universe. Assuming the inflationary paradigm, this means that these density perturbations are linked to the details of the adopted inflationary model. The simplest model of inflation, a single slowly-rolling scalar field, predicts nearly Gaussian initial density perturbations, while more complicated models lead to non-Gaussianities of different shapes, amplitudes and scale-dependences.

If the initial density perturbations are purely Gaussian, the density field can be fully described by the two-point correlation function (i.e. the power spectrum in Fourier space), since all higher order moments are zero. The three-point correlation function (i.e. the `bispectrum' in Fourier space) is the lowest order statistic affected by the presence of non-Gaussianities. It can be measured by considering triangles in the Fourier space, and the shape of the triangles, i.e. the relative magnitude of the wave vectors, is linked to the physical mechanism creating such non-Gaussianities.

In this work, we consider non-Gaussianities of `local' type, corresponding to `squeezed' triangles in which two wave vectors have similar magnitudes and are much larger than the third one ($ k_1 \simeq k_2 \gg k_3$). This type of primordial non-Gaussianity originates from inflationary models whose non-linearities develop on super-horizon scales (e.g. `curvaton' models, multi-field inflation, see \citealt{Yadav2010}). 

To compute the impact of primordial non-Gaussianities on early structure formation, we firstly generate the non-Gaussian initial conditions at $z~=~200$ using a parallel non-Gaussian code developed and updated in \citet{Nishimichi2009,Valageas2011,Nishimichi2012}. This code is based on second-order Lagrangian perturbation theory \citep[e.g.,][]{Scoccimarro1998,Crocce2006} with pre-initial particles placed on a regular lattice. We consider the phenomenological model for scale-dependent non-Gaussianities of \cite{Becker2011} (i.e. {\it generalized local ansatz})
\begin{equation} 
\zeta(\mathbf{x}) = \zeta_\rmn{G}(\mathbf{x}) + \frac{3}{5}\left[\fNL * (\zeta_\rmn{G}^2 - \langle{\zeta_\rmn{G}^2}\rangle)\right](\mathbf{x}) \, ,
\label{eq:localansatz}
\end{equation}
where $\zeta$ denotes the curvature perturbation, $\zeta_\rmn{G}$ a Gaussian random field and \fNL\ the scale-dependent amplitude of non-Gaussianity. The convolution is defined in Fourier space as a multiplication by a scale-dependent coefficient
\begin{equation} 
\fNL (k) = \fNLn \,\left(\frac{k}{k_0}\right)^{\gamma} \, ,
\label{eq:nGmodel}
\end{equation}
where \fNLn\ indicates the magnitude of non-Gaussianity at the scale $k=k_0$.

\begin{figure}
	\centering
	\resizebox{\hsize}{!}{\includegraphics{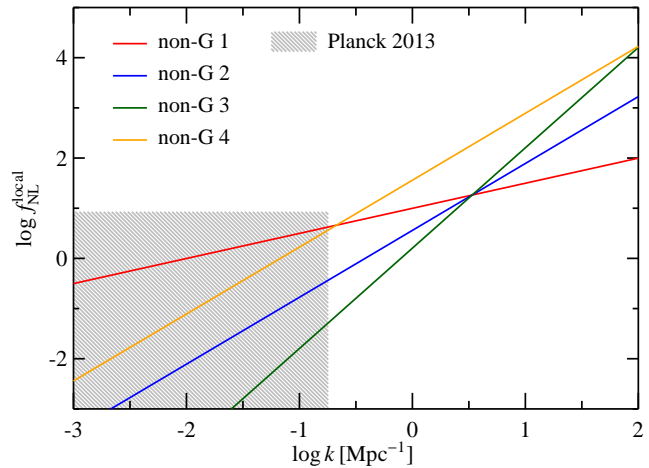}}
	\caption{Level of non-Gaussianity of local type at different scales for the different models adopted in this work, parametrized with a power law as in Equation~\ref{eq:nGmodel}. The red line indicates the non-Gaussian model 1 ($\fNLn=82$ and $\gamma=1/2$); blue the non-Gaussian model 2 ($\fNLn=10^3$ and $\gamma=4/3$); dark-green the non-Gaussian model 3 ($\fNLn=7357$ and $\gamma=2$) and orange the non-Gaussian model 4 ($\fNLn=10^4$ and $\gamma=4/3$). The hatched region shows the allowed region of \fNLn\ at scales $k \lesssim 0.18 \,\MpcInv$ (68 \% credible interval) from \citet{Planck2013_24}.}
	\label{fig:fNL}
\end{figure}

This model is a generalization of the so-called local-type non-Gaussianity \citep{Komatsu2001}. It has two parameters, \fNLn\ and $\gamma$, that determine the amplitude and the slope of the function \fNL(k), respectively. We recover the (scale-independent) local non-Gaussianity by setting $\gamma=0$. The parameters of the models used in this work are taken from paper~I, listed in Table~\ref{tab:simul} and plotted in Fig.~\ref{fig:fNL}. We also show in Fig.~\ref{fig:fNL} the constraint on \fNL(k) by \citet{Planck2013_24} as a gray shaded region. We consider models with $\gamma>0$ such that the magnitude of non-Gaussianity is large on galactic scales, while being below the observational upper limit on the scales probed by \Planck\ observations. Also, the sign convention for \fNL\ is chosen such that a positive \fNL\ leads to a positive skewness  in the initial matter fluctuations. 

 \subsection{Dark matter simulations}

Previous studies have shown, analytically and by means of N-body simulations, the effect of scale-invariant primordial non-Gaussianities on the halo mass function and halo power spectrum \citep{Nishimichi2010,Nishimichi2012}. In this work. we appeal to the N-body simulations presented in paper~I, which we briefly summarize below, to study how scale-dependent initial non-Gaussianities affect structure formation at redshift $7 \le z \le 15$.

\begin{table}
	\centering
	\begin{tabular}{c c c c}
\toprule

\textbf{Simulation }	         & \textbf{$\boldsymbol{\fNLn}$} & $\boldsymbol{\gamma}$ & \textbf{Haloes at $\boldsymbol{z=6.5}$ } \\

\midrule

\vspace{\colhspace}
Gaussian	            				&    -  								&  - 		& \np{331139} \\
\vspace{\colhspace}
non-Gaussian 1				&      82 							& 1/2  	&  \np{324181} \\
\vspace{\colhspace}
non-Gaussian 2				&      $10^3$ 					& 4/3 	&  \np{337388} \\
\vspace{\colhspace}
non-Gaussian 3				&      $7.357\times10^3$	 & 2 		&  \np{374653} \\
\vspace{\colhspace}
non-Gaussian 4				&      $10^4$						 & 4/3 	&  \np{491577} \\

\bottomrule
	\end{tabular}
	\caption{Parameters defining the non-Gaussian term in the initial density perturbations (see Equation~\ref{eq:nGmodel}) at the pivot wave number $k_0 = 100\,h\,\mathrm{Mpc}^{-1}$, and number of haloes (excluding satellites) identified by \adaptahop\ at the final step of our simulations ($z=6.5$).}
	\label{tab:simul}	
\end{table}

We adopt the smoothed particle hydrodynamics code \gadget\ \citep{Springel2001,Springel2005} to run the N-body simulations. We fix the cosmological parameters to the values obtained by the \Planck\ mission \citep[see last column `\Planck $+$WP$+$highL$+$BAO' in Table 5 of][]{Planck2013_16}, namely $\Omegal = 0.693$,  $\Omegam = 0.307$, $h = 0.678$ and $\sigma_8=0.829$.
We perform all the simulations in a periodic box of side $50 \, \hInvMpc$ with $1024^3$ particles; the mass of each dark matter particle is therefore $\sim 9.9 \times 10^6 \, \hInvMsun $ 

We run a set of 5 different N-body simulations, starting each simulations at  $z=200$ and evolving it till $z=6.5$, taking snapshots every $\sim 40 \, \Myr$. We identify dark matter haloes and subhaloes by means of \adaptahop\ \citep{Aubert2004}, a (sub)structure finder based on the identification of saddle points in the (smoothed) density field, fixing the density threshold $\delta \rho / \rho_c = 180$, where $\rho_c$ is the critical density of the Universe. We consider only haloes containing $> 20$ particles ($\Mhalo \ga 2.9 \times 10^8 \, \Msun$) and build the merger trees by means of \treemaker\ \citep{Tweed2009}.

\subsection{The~\citet{Mutch2013} galaxy formation model}\label{sec:Mutch}

Many approaches have been adopted to describe the assembly of stellar mass in dark-matter haloes. Semi-analytic models \citep[e.g.][]{Kauffmann1993,Cole2000,Croton2006,Lu2011} are multi-parameters models which adopt  analytic relations to describe the baryonic processes (e.g. gas accretion, galaxy interactions, stellar and AGN feedback) that regulate star formation and chemical enrichment in galaxies. Halo occupation distribution models (HOD) are defined in terms of the statistical properties of dark-matter haloes, i.e. of the probability of a halo of given mass to host a given number of galaxies, without any explicit connection to the physical processes acting in galaxies \citep[e.g.][]{Berlind2002,Berlind2003,Zheng2005}. Halo abundance matching (HAM) is another statistical approach in which the cumulative mass functions of haloes and galaxies are matched under the assumption of a monotonic relation between stellar and halo masses (i.e. more massive haloes contains more massive galaxies) \citep[e.g.][]{Conroy2006,Moster2013}. 

In spite of the variety of these approaches, they all suffer from limitations which do not make them suitable for our purposes. Semi-analytic models suffer from the presence of many adjustable parameters (usually $\gtrsim 10$) that can be partially constrained by observations in the local Universe \citep[e.g.][]{Lu2011,Henriques2013}, but remain largely unconstrained at higher redshift. Statistical approaches such as HOD and HAM consider only the average properties of haloes of given mass, therefore not accounting for the stochasticity that is inherent to the hierarchical growth of dark matter haloes. 

These reasons have motivated over the last years the emergence of another kind of galaxy formation model in which the complexity of the baryonic physics is subsumed and simplified into a few analytic functions \citep[e.g.][]{Tacchella2013,Behroozi2013,Mutch2013}. The advantage of these phenomenological models is that they do not suffer from the presence of multiple, unconstrained parameters typical of semi-analytic models, while allowing one to link the evolution of stellar mass in dark matter haloes to physical quantities, and not just to the average statistical properties of haloes. 

Among the various available phenomenological models, we adopt the galaxy formation model of \citet{Mutch2013} (hereafter M13), since it allows us to compute the stellar mass assembly associated with the hierarchical growth of {\em individual} dark-matter haloes, hence accounting for the stochasticity of stellar mass growth in galaxies (e.g. see fig~1 of M13). In practice, the model of M13 relates the stellar mass growth in dark-matter haloes to two analytic functions: a `growth' function, describing the amount of baryons available for star formation, and a `physics' function, which determines the fraction of available baryons actually converted into stars. The variation of the stellar mass of a halo is therefore described by the following relation
\begin{equation}\label{eq:dM}
\frac{d\Mstar}{dt} =  \Fgrowth \, \Fphys \, ,
\end{equation} 
where the growth function \Fgrowth\ is defined as
\begin{equation}\label{eq:Fgrowth}
\Fgrowth = \fb\frac{d\Mvir}{dt} \, ,
\end{equation}
where $f_\txn{b}=0.17$ is the universal fraction of baryons. The previous relation implies that the rate of change in the amount of available baryons is proportional to the rate of change in the virial mass of the halo, with the constant of proportionality being \fb. 

We adopt a log-Cauchy function of halo mass to describe the `physics' function, unlike M13 who adopt a log-normal function. The reason is that our study focuses on haloes of lower mass than those considered by M13, and the log-normal function proposed by M13 decays very rapidly far from the peak, implying an unphysical low efficiency of star formation at low halo masses. We therefore describe the `physics' function as
\begin{equation}\label{eq:Fphys}
\Fphys =  \epsilon \, \frac{\sigma^2}{\left ( \log\Mvir-\log\MvirP \right ) ^2 + \sigma^2} \, ,
\end{equation} 
and fix $\log (\MvirP/\Msun) = 11.8$, $\sigma = 1$ and $\epsilon = 0.07$, all independent of redshift, to match the observed far-UV luminosity function of \citet{Bouwens2014} at $z=7$ (see Section~\ref{sec:UV_LF}). Equation~\ref{eq:Fphys} implies an efficiency of baryons conversion into stars $\sim 0.04$ at $\log (\Mvir/\Msun) = 11$, and $\sim 0.008$ at $\log (\Mvir/\Msun) = 9$.

Equations~\eqref{eq:dM}-\eqref{eq:Fphys} allow us to compute the star formation history from redshift 20 to 6.5 for all haloes identified in the Gaussian and non-Gaussian simulations. However, the M13 model does not include any prescription for the chemical evolution of galaxies, as they assume a constant value for the metallicity, $1/3$ of the solar value, at all ages and galaxy masses. Unlike M13, we associate a value for the metallicity $\FeH(t)$ drawn from the mass-metallicity relation for dwarf galaxies of \citet{Kirby2013} (see their Eq. 4). 
\begin{equation}\label{eq:Kirby}
\FeH(t) = -1.69 + 0.3 [ \log(\Mstar/\Msun)-6 ] \, ,
\end{equation}
assuming solar-scaled abundance rations and a dispersion of 0.17 dex around this mean relation. This relation implies a metallicity $\FeH = -1.54$ at $\log(\Mstar/\Msun)=6.5$ and   $\FeH = -0.79$ at $\log(\Mstar/\Msun)=9$.

\subsection{Spectral evolution model}

To compute the light emission from galaxies, we combine the star formation and chemical enrichment histories obtained with the M13 model with the population synthesis code of \citet{BC03}. We adopt the Galactic-disc stellar initial mass function of \citet{Chabrier2003}, with lower and higher mass cut-offs at 0.1 and 100 \Msun, respectively.\footnote{The catalogues of halo masses, galaxy stellar masses and spectral energy distributions computed from the different Gaussian and non-Gaussian simulations in the redshift range $15 \le z \le 7$ are available upon request from the corresponding author.}

\section{Results}\label{sec:results}

In the previous section, we have described the different N-body simulation we have run with Gaussian and non-Gaussian initial conditions, and the tools we have used to identify the dark matter haloes and build the merger trees. We have also described the M13 galaxy formation model, which we have used to compute the assembly of stellar mass with time in the dark matter haloes previously identified. Finally, we have introduced the BC03 population synthesis code, which we adopted to calculate the spectral energy distribution of each galaxy in the catalogues. With these tools we can compute the halo and stellar mass functions, and the far-UV luminosity function at different redshifts and for the different N-body simulation. In this section, we will show the results of these computations, showing the redshift evolution of these quantities for the Gaussian simulation, and the differences introduced by the primordial non-Gaussianities as a function of redshift and halo mass. Results on the halo and stellar mass functions, and associated constraints, have been previously presented in Paper I, although assuming a different galaxy formation model. 

\subsection{Halo mass function}\label{sec:haloMF}

\begin{figure}
	\centering
	\resizebox{\hsize}{!}{\includegraphics{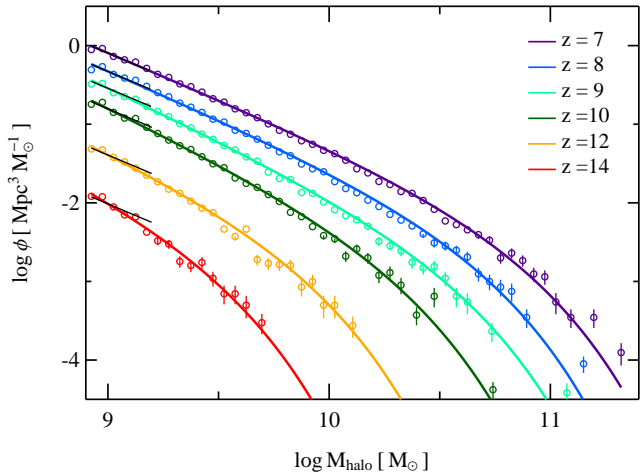}}
	\caption{Halo mass function at different redshifts for the simulation with Gaussian initial conditions. Open circles indicate the mass function measured from our halo catalogue, while solid lines are a Schechter function fit to the points. Solid black lines are power laws $\Phi(\M) \propto 10^{(\M-\M^\star)(1+\alpha)}$ with $\alpha$ equal to the value obtained at $z=7$ and scaled to match the computed mass functions at $\log (\Mhalo/\Msun) = 9 $, and are meant to highlight the evolution of the low-mass slope of the halo mass function with redshift. Error-bars are computed assuming independent Poisson distributions in each mass bin, and are plotted only when greater than the size of circles.}
	\label{fig:Halo_MF_G}
\end{figure} 

We show in Fig.~\ref{fig:Halo_MF_G} the evolution with redshift of the halo mass function obtained from the simulation with Gaussian initial conditions (see also paper~I). The open circles of different colours represent the halo mass function at different redshifts measured from our simulation. To compute the halo mass function, we divide the halo catalogue in logarithmic mass bins of 0.05 dex width in the halo mass range $8.9\le \log (\Mhalo/\Msun) \le 11.4$, and then consider the number of haloes within each bin and divide this number by the simulation volume and bin width.\footnote{We do not include satellite haloes in the computation of the halo mass function and in the successive computations. However, we note that the number of satellites at these redshifts is small, so the choice of including/neglecting them would not influence our results.} The limited volume of our simulations makes massive haloes rare, therefore increasing the Poisson noise at large halo masses. To reduce this noise, we merge adjacent bins until reaching $\ge 10$ objects in each bin. Error-bars are computed assuming Poisson noise.
The solid lines in Fig.~\ref{fig:Halo_MF_G} are obtained by fitting the binned luminosity function with a \citet{Schechter1976} function of the form
\begin{equation}\label{eq:Schechter_MF}
\Phi(\M) = \ln(10) \, \Phi^\star \, 10^{(\M-\M^\star)(1+\alpha)} \, \exp \left( -10^{\M-\M^\star}\right) \, ,
\end{equation}
where $\M = \log(\Mhalo/\Msun)$, $\alpha$ is the slope of the power law at low masses, $\M^\star = \log(\txn{M}_\txn{halo}^\star/\Msun)$ is the characteristic mass and $\Phi^\star$ the normalisation. We adopt a Markov Chain Monte Carlo to find the combination of parameters $\M^\star$, $\alpha$ and $\Phi^\star$ that best reproduce our measurements, choosing as best-fit values the median of the posterior marginal distribution of each parameter. The adoption of an MCMC also allows us to derive reliable estimates on the uncertainties of the Schechter function parameters.
To ease the comparison of the low-mass slope of the halo mass function we also over-plot with black lines in Fig.~\ref{fig:Halo_MF_G} a power law of the form $\Phi(\M) \propto 10^{(\M-\M^\star)(1+\alpha)}$ fixing $\alpha$ to the value obtained with the Schechter function fit at redshift $z=7$ and normalising the function to match the measured mass function at $\log(\Mhalo/\Msun) = 9$.

Fig.~\ref{fig:Halo_MF_G} indicates that the number density of haloes decreases with increasing redshift, at fixed halo mass, as expected in the hierarchical growth of structures in a \LCDM\ Universe. The characteristic halo mass also decreases with increasing redshift, from $\M^\star=10.78\pm0.05$ at $z=7$ to $\M^\star=9.39\pm0.12$ at $z=14$. The low-mass slope of the halo mass function also evolves with redshift, becoming steeper with increasing redshift, from $\alpha=-2.19\pm0.02$ at $z=7$ to $\alpha=-2.35\pm0.1$ at $z=10$. At higher redshift, the smaller values of the characteristic halo mass $\M^\star$ combined with the resolution of our simulation do not allow us to properly measure the slope of the low-mass end of the halo mass function, as the exponential cut-off of the mass function makes the fitted value $\alpha$ unreliable. 
The behaviour obtained for the halo mass function is expected in a \LCDM\ Universe in which structures grow hierarchically from low to high masses, since merging increases, as the age of the Universe increases, the relative number of high mass haloes with respect to the low mass ones, thereby  flattening the low-mass end of the mass function, while increasing the characteristic mass and number density of haloes.

\begin{figure*}
	\centering
	\resizebox{\hsize}{!}{\includegraphics{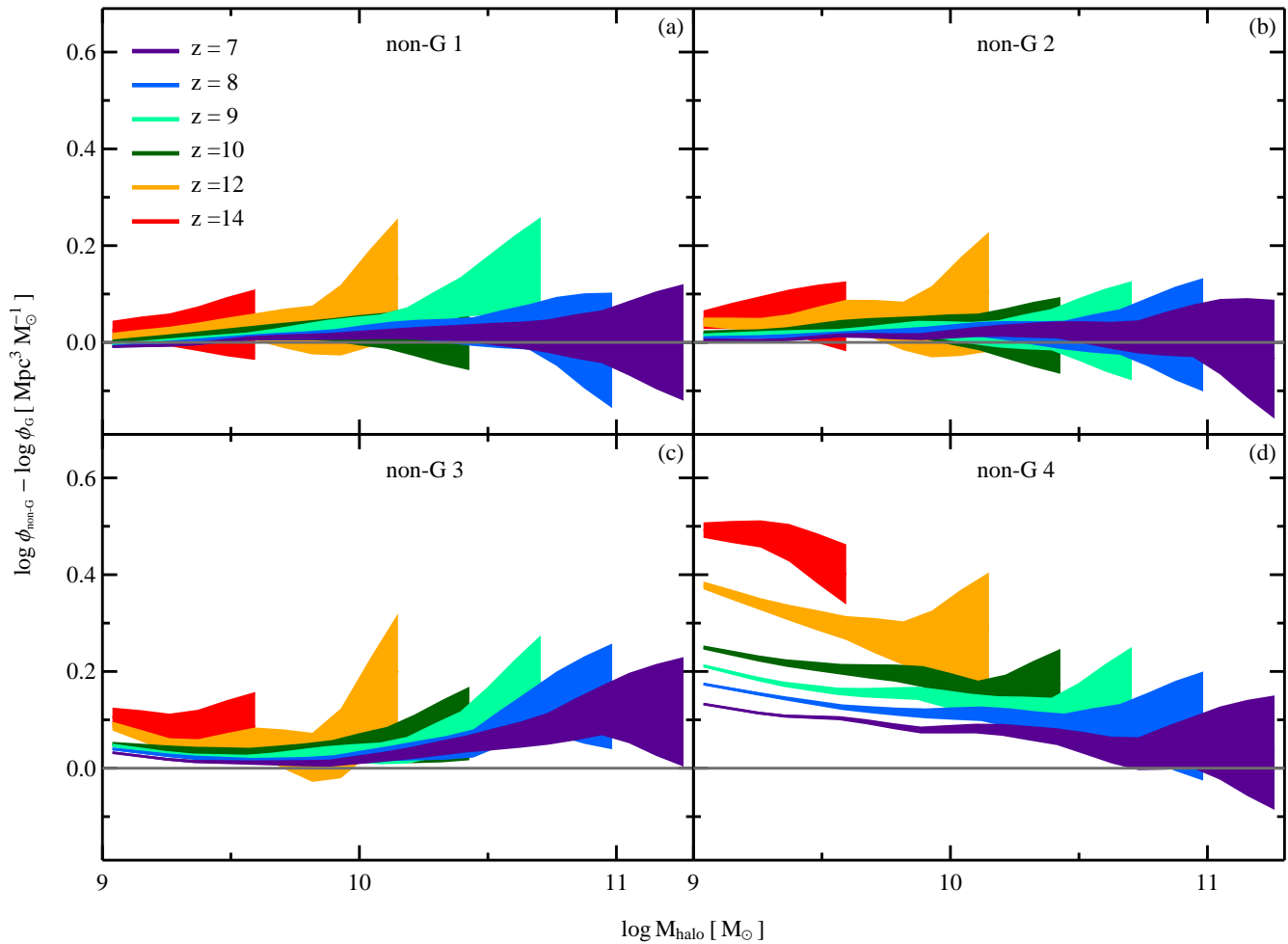}}
	\caption{Impact of primordial non-Gaussianities on the halo mass function. (a) Difference between the halo mass function computed from the non-Gaussian simulation 1 and that computed from the Gaussian simulation, at redshift $z=7$ (purple region), 8 (blue), 9 (light green), 10 (dark green), 12 (orange), 14 (red). (b) Same as (a), but for the non-Gaussian simulation 2. (c) Same as (a), but for the non-Gaussian simulation 3. (d) Same as (a), but for the non-Gaussian simulation 4. Each coloured region is computed assuming independent Poisson errors for each simulation and mass bin (see Section~\ref{sec:haloMF}).}
	\label{fig:Halo_MF}
\end{figure*}

We show in Fig.~\ref{fig:Halo_MF} the difference between the halo log-mass function computed from each simulation with non-Gaussian initial density perturbations and the simulation with Gaussian perturbations as a function of halo mass, for different redshifts (see also paper~I). To accomplish this, we divide, at each redshift, the catalogue of halo masses of each simulation in 10 logarithmic bins in the range $8.9\le \log (\Mhalo/\Msun) \le 11.4$. We then divide the number of haloes in each bin by the simulation volume and bin width, obtaining the halo number density, and compute the difference $\delta\phi=(\log \phi_\txn{non-G}-\log \phi_\txn{G})$ between the halo number density for each non-Gaussian simulation and the Gaussian one as a function of halo mass. We only consider bins with $\ge20$ objects, and calculate the error in each bin by summing in quadrature the relative errors in the halo number density for the Gaussian and non-Gaussian simulation, which are in turn computed assuming a Poisson distribution.     
The different colours in Fig.~\ref{fig:Halo_MF} refer to different redshifts, and the width of the coloured regions reflects the error associated to each bin. 

Fig.~\ref{fig:Halo_MF}(a) indicates that the shallow spectrum ($\gamma=1/2$), with a low normalisation ($\fNLn=82$), used to generate the non-Gaussian initial conditions for the simulation non-G~1 produces a halo mass function which is statistically consistent with that obtained from the Gaussian simulation, at all redshift and halo masses. 

Fig.~\ref{fig:Halo_MF}(b) indicates that the steeper slope ($\gamma=4/3$) and larger normalization ($\fNLn=10^3$) adopted to generate the initial conditions for the non-G~2 simulation produces marginally larger differences in the halo mass function than those obtained for the non-G~1 model. These differences are $<0.1\,\txn{dex}$ at all redshift and halo masses here considered, and are more significant at low halo masses ($\log (\Mhalo/\Msun) \lesssim 9.5$) because of the higher number of low- than high-mass haloes. We note also that the effect of initial non-Gaussianities increases with increasing redshift, at fixed halo mass. 

Fig.~\ref{fig:Halo_MF}(c) shows that the non-G~3 model ($\gamma=2$ and $\fNLn=7.357\times10^3$), which has stronger initial non-Gaussianities then the non-G~1 and 2 models, produces statistically significant variations in the halo mass function (up to 0.2 dex) with respect to the Gaussian simulation, at all redshifts and both at low and high masses. At fixed redshift, the non-G~3 halo mass function deviates the most from the Gaussian mass function at the extreme of the mass range, while being more similar at $\log (\Mhalo/\Msun) \sim 10$; i.e. the effect of the primordial non-Gaussianities in this model is to increase the number of both low- and high-mass haloes. As for the non-G~2 simulation, these deviations become more statistically significant at low halo masses because of the larger number of low- than high-mass haloes in our simulation volume. We note also that at fixed halo mass the difference between the non-G~3 and Gaussian halo mass function increases with increasing redshift, following the same qualitative trend observed for the non-G~2 simulation.

Fig.~\ref{fig:Halo_MF}(d) indicates that the halo mass function of the non-G~4 model ($\gamma=4/3$ and $\fNLn=10^4$) exhibits differences up to $\sim 0.5$ dex with respect to that of the Gaussian simulation. These differences are much larger, at all redshift and halo masses, than those observed for the non-G~1, non-G~2 and non-G~3 simulations. This behaviour is expected, as the non-G~4 simulation has the strongest level of primordial non-Gaussianities at scales $\log (k/\txn{Mpc}) < 2$ with respect to all other non-Gaussian models adopted in this work (see Fig.~\ref{fig:fNL}). At fixed mass, the number of low-mass haloes increases with increasing redshift with respect to the Gaussian simulation, in a consistent manner to what obtained for the non-G~1, non-G~2 and non-G~3 simulations. At fixed redshift, the difference in the number density of haloes  with respect to the Gaussian simulation increases with decreasing halo mass, unlike the results from the non-G~3 simulation. 

The trends shown in Fig.~\ref{fig:Halo_MF} for the different non-Gaussian models can be understood as follows. First of all, a (scale-independent) positive skewness in the matter fluctuations boosts the formation of rare objects, i.e. massive halos, since it increases their probability to cross the threshold density contrast required for collapse \citep[e.g.][]{Matarrese2000}. However, adopting a scale-dependent skewness complicates this picture, as for `blue' tilts, i.e. a skewness that increases with decreasing scale, the formation of low mass halos is also amplified. The reason is that in these models the skewness is larger on smaller scales, those corresponding to small-scale fluctuations, i.e. to low mass halos. The net effect of a scale-dependent primordial non-Gaussianity with a `blue' spectrum is therefore the combination of these two effects: a boost of rare fluctuations (i.e. massive haloes), \emph{and} an increasingly stronger boost of fluctuations with decreasing scale, i.e. decreasing halo mass, with the relative strengths of these effects being determined by the shape and normalisation of the adopted model for the non-Gaussianity.

In the next section we will show how the effect of primordial non-Gaussianities on the number density of haloes at different redshifts affects the redshift evolution of the galaxy stellar mass function.

\subsection{Galaxy stellar mass function}\label{sec:galMF}

\begin{figure}
	\centering
	\resizebox{\hsize}{!}{\includegraphics{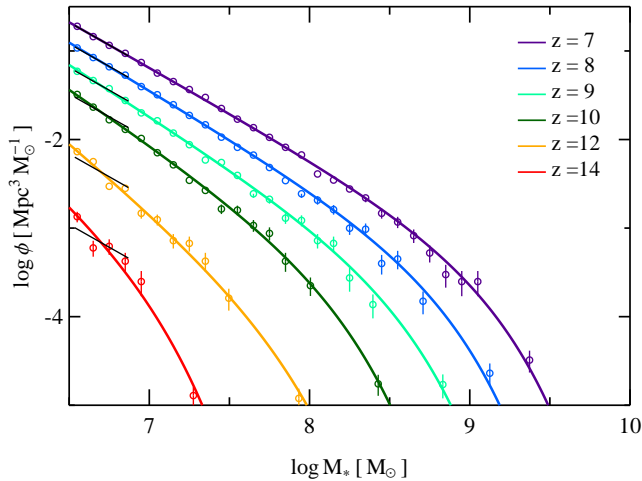}}
	\caption{Galaxy stellar mass function at different redshifts for the simulation with Gaussian initial conditions. Open circles indicate the stellar mass function computed form our simulation, while solid coloured lines are fit to the points with a Schechter function. Solid black lines are power laws $\Phi(\M) \propto 10^{(\M-\M^\star)(1+\alpha)}$ with $\alpha$ equal to the value obtained at $z=7$ and scaled to match the computed mass functions at $\log (\Mstar/\Msun) = 6.7 $. Error-bars are computed assuming independent Poisson distributions in each mass bin, and are plotted only when greater than the size of circles.}
	\label{fig:Gal_MF_G}
\end{figure}

We have shown that the adoption of non-Gaussian initial density perturbations produces measurable differences in the halo mass function for the non-Gaussian simulations 2, 3 and 4, i.e. for those simulations with strong enough initial non-Gaussianities. We therefore expect a similar behaviour for the galaxy stellar mass function, as it depends on the halo merger history (see equations~\eqref{eq:dM}-\eqref{eq:Fphys}).

We show in Fig.~\ref{fig:Gal_MF_G} the galaxy stellar mass function obtained by applying the galaxy formation model of M13 (see Section~\ref{sec:Mutch}) on the dark-matter simulation with Gaussian initial conditions. The open circles of different colours indicate the mass function at different redshifts. We compute the galaxy stellar mass function in the same way as we do for the halo mass function (see Section~\ref{sec:haloMF}), but adopting logarithmic mass bins of 0.1 dex width in the range $6.5\le \log (\Mstar/\Msun) \le 10$. The solid, coloured lines in Fig.~\ref{fig:Gal_MF_G} are obtained by fitting a Schechter function to the our measured points (see equation~\eqref{eq:Schechter_MF} and Sec.~\ref{sec:haloMF} for details). The solid black lines in Fig.~\ref{fig:Gal_MF_G} are power laws of the form $\Phi(\M) \propto 10^{(\M-\M^\star)(1+\alpha)}$ with $\alpha$ equal to the value obtained with the Schechter function fit at redshift $z=7$, and normalised to match the measured stellar mass function at  $\log(\Mstar/\Msun) = 6.7$.

Fig.~\ref{fig:Gal_MF_G} shows that, as for the halo mass function of Fig.~\ref{fig:Halo_MF_G}, the number density of galaxies decreases with increasing redshift, at fixed stellar mass. The characteristic stellar mass also decreases with increasing redshift, from $\M^\star=9.03\pm0.12$ at $z=7$ to $\M^\star=6.90\pm0.36$ at $z=14$, indicating that the typical stellar mass of galaxies decreases with increasing redshift. The low-mass end of the mass function steepens with increasing redshift, from $\alpha=-2.03\pm0.02$ at $z=7$ to $\alpha=-2.23\pm0.07$ at $z=10$. As we already noted for the halo mass function, at higher redshift the low values of the characteristic stellar mass combined with the resolution of the N-body simulation makes the exponential cut-off important at all stellar masses here considered, therefore affecting the value of $\alpha$ obtained with the Schechter function fit. The likely explanation for the flattening of the low-mass end of the stellar mass function from high to low redshift is merging, as we already noted for the halo mass function.
We note also that at a given redshift the low-mass slope of the stellar mass function is flatter than the halo mass function ($\alpha_\txn{halo}-\alpha_\txn{star} \sim 0.15$ at $z=7$--10). This is due to the increasing efficiency of baryon conversion into stars with increasing halo mass in the galaxy formation model adopted in this work, for the halo mass ranges here considered (see Section~\ref{sec:Mutch} and equation~\ref{eq:Fphys}). This causes high mass haloes to have, on average, larger stellar-to-halo mass ratios ($\Mstar/\Mhalo$) than haloes with lower mass, thus increasing the relative number of high- to low-mass galaxies, i.e. flattening the stellar mass function with respect to the halo mass function. 

\begin{figure*}
	\centering
	\resizebox{\hsize}{!}{\includegraphics{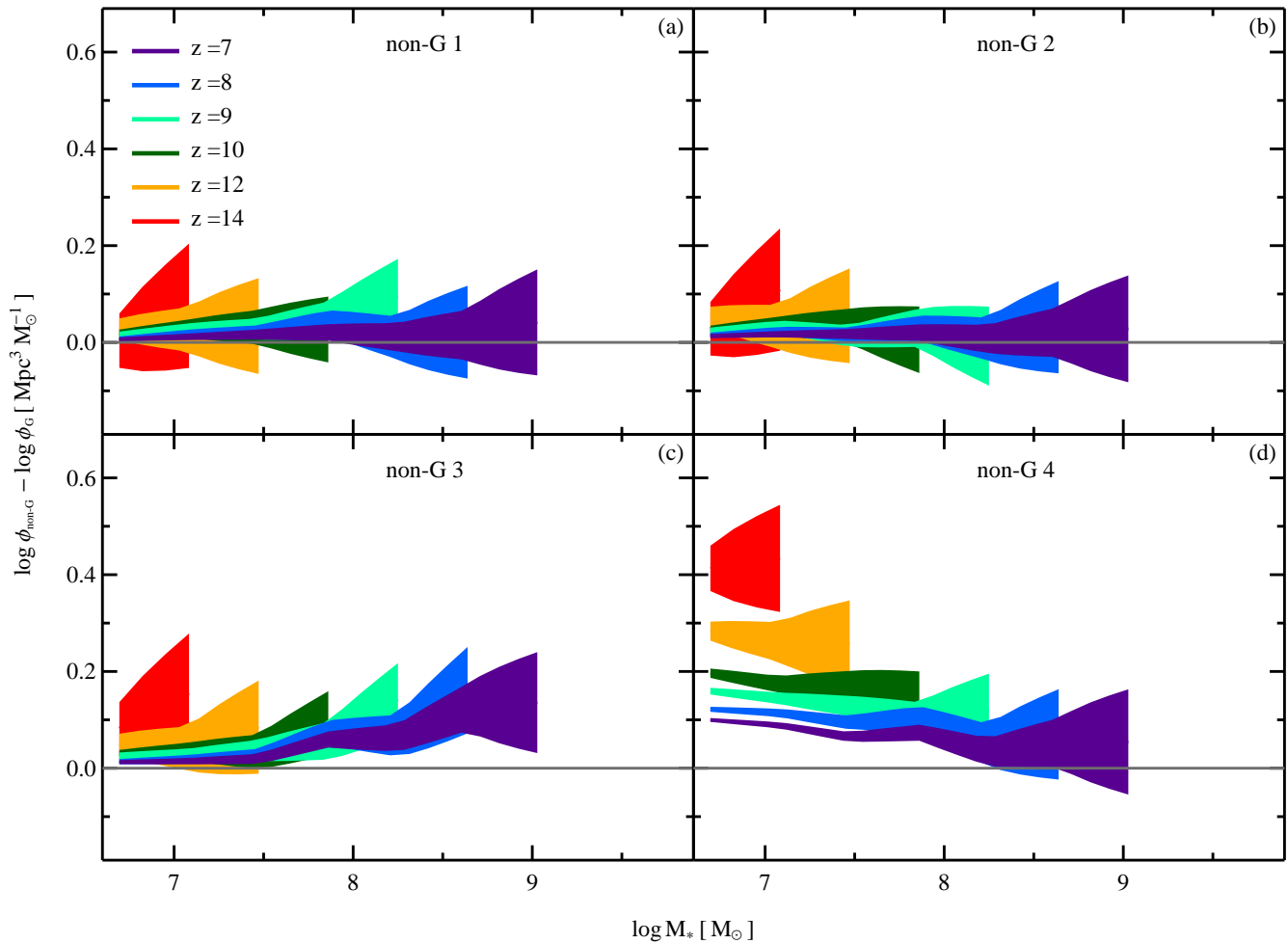}}
	\caption{Impact of primordial non-Gaussianities on the galaxy stellar mass function. (a) Difference between the galaxy stellar mass function computed from the non-Gaussian simulation 1 and that computed from the Gaussian simulation, at redshift $z=7$ (purple region), 8 (blue), 9 (light green), 10 (dark green), 12 (orange), 14 (red). (b) Same as (a), but for the non-Gaussian simulation 2. (c) Same as (a), but for the non-Gaussian simulation 3. (d) Same as (a), but for the non-Gaussian simulation 4. Each coloured region is computed assuming independent Poisson errors for each simulation and mass bin.}
	\label{fig:Gal_MF}
\end{figure*} 

We show in Fig.~\ref{fig:Gal_MF} the effect of non-Gaussian initial density perturbations on the galaxy stellar mass function as a function of stellar mass. We compute the stellar mass function for the Gaussian simulation and each non-Gaussian simulation in the same way as we do for the halo mass function (see Section~\ref{sec:haloMF}), adopting 10 logarithmic bins in the range $6.5\le \log (\Mstar/\Msun) \le 10$, and sconsidering only bins with $\ge20$ objects. %We then calculate, at each redshift, the difference between the number density of galaxies obtained from each non-Gaussian simulation and the Gaussian one as a function of stellar mass. The different colours in Fig.~\ref{fig:Gal_MF} correspond to different redshifts, and the width of the coloured regions indicates the error in each bin, which we compute by summing in quadrature the relative errors in the galaxy number density for the Gaussian and non-Gaussian simulation, which are in turn computed assuming a Poisson distribution.        

Fig.~\ref{fig:Gal_MF}(a) displays the difference between the galaxy stellar mass function computed from the non-G~1 simulation and that computed from the Gaussian simulation. As for the halo mass function [see Fig.~\ref{fig:Halo_MF}(a)], the stellar mass function obtained from the non-G~1 model is consistent, within the errors, with that computed from the Gaussian simulation, at all redshift and masses. 

Fig.~\ref{fig:Gal_MF}(b) indicates that, unlike the galaxy stellar mass function of the non-G~1 simulation, that of the non-G~2 simulation exhibits, at low ($\log(\Mstar/\Msun)\lesssim 8$) masses, small ($<0.1$ dex) but statistically significant differences with the mass function of the Gaussian simulation. In particular, the number of low-mass galaxies is increased with respect to the Gaussian simulation. Also, this effect increases with increasing redshift, at fixed stellar mass, as already noted for the halo mass function (Section~\ref{sec:haloMF}).

Fig.~\ref{fig:Gal_MF}(c) shows that the primordial non-Gaussianities adopted in the non-G~3 simulation, stronger than those in the non-G~1 and non-G~2 simulations, produce even more statistically significant differences with respect to the stellar mass function of the Gaussian simulation, both at low and high stellar masses. At each redshift, the number density of low- and high-mass galaxies is increased with respect to the Gaussian simulation, but, unlike the corresponding figure for the halo mass function [Fig.~\ref{fig:Halo_MF}(c)], the effect is more pronounced for high-mass galaxies. We can explain this behaviour by appealing to the decreasing efficiency of baryon conversion into stars, i.e. the decrease of the stellar-to-halo mass ratio, with decreasing halo mass, which reduces the differences between the non-G~3 and Gaussian simulations at low stellar masses. 
At fixed stellar mass, the effect of primordial non-Gaussianities on the galaxy stellar function increases with increasing redshift, as we already noted for the non-G~2 simulation.

Fig.~\ref{fig:Gal_MF}(d) indicates that the non-G~4 simulation, which has the strongest level of primordial non-Gaussianities among all simulations, produces the strongest deviations in the galaxy stellar mass function with respect to the Gaussian simulation, as already noted for the halo mass function in Fig.~\ref{fig:Halo_MF}(d). At all redshift and masses the number density of galaxies is larger in the non-G~4 simulation than in the Gaussian simulation. This difference increases, at fixed stellar mass, with increasing redshift, reaching $\delta \phi \sim 0.4$ at $z=14$ and $\log(\Mstar/\Msun) = 6.5$. We note that this value is lower than that observed in the halo mass function [$\delta \phi \sim 0.5$, see Fig.~\ref{fig:Halo_MF}(d)], and that, even though at fixed redshift the difference between the non-Gaussian and Gaussian mass functions increases with decreasing galaxy stellar mass, this trend is weaker (i.e. flatter) that that for the halo mass function. This is in qualitative agreement to what we noted for the non-G~3 simulation and likely related to the same explanation: the decreasing baryon conversions efficiency with decreasing halo mass reduces the differences between the non-Gaussian simulations and the Gaussian one at low stellar masses. 

Our results on the effects of non-Gaussianity on the galaxy stellar mass function are qualitatively very similar to those found in paper~I, despite the different galaxy formation models used \citep[see fig 3 of][]{Habouzit2014}.

In the next section, we will show by means of the \citet{BC03} population synthesis code how the differences in the galaxy stellar mass function between the non-Gaussian and Gaussian simulations reflect into different far-UV luminosity functions at different redshifts.

\subsection{UV luminosity function}\label{sec:UV_LF}

\begin{figure}
	\centering
	\resizebox{\hsize}{!}{\includegraphics{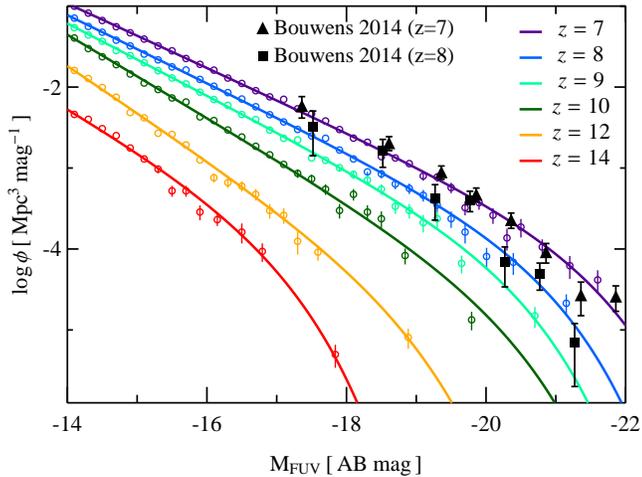}}
	\caption{Far-UV luminosity function at different redshifts for the simulation with Gaussian initial conditions. Open circles indicate the far-UV luminosity function computed form our simulation, while solid coloured lines are fit to the points with a Schechter function. Error-bars are computed assuming independent Poisson distributions in each mass bin, and are plotted only when greater than the size of circles. Black-filled symbols indicate the UV luminosity function measured by \citet{Bouwens2014} at $z\sim7$ (598 galaxies, triangles) and $z=8$ (225 galaxies, squares).}
	\label{fig:UV_LF_G}
\end{figure}

In the previous sections we have illustrated the effect of primordial non-Gaussianities on the halo and stellar mass functions computed from our N-body simulation and the M13 galaxy formation model. By means of the galaxy chemical evolution prescription presented in Section~\ref{sec:Mutch} (see Equation~\ref{eq:Kirby}) and the \citet{BC03} population synthesis code, we now show the effect of non-Gaussian initial density perturbations on the redshift evolution of the far-UV luminosity function.

Fig.~\ref{fig:UV_LF_G} shows the far-UV luminosity function computed from the Gaussian simulation. The open circles of different colours indicate the far-UV luminosity function at different redshifts, which we compute in the same way as we do for the halo mass function (see Section~\ref{sec:haloMF}), but adopting bins of far-UV absolute magnitudes 0.2 dex width in the range $-22 \le \Muv \le -14 $. The solid, coloured lines in Fig.~\ref{fig:UV_LF_G} are computed by fitting with the same method outlined in Sec.~\ref{sec:haloMF} a Schechter function of the form 
\begin{equation}\label{eq:Schechter_LF}
\phi(M) = 0.4 \, \ln(10) \, \phi^\star \, 10^{-0.4(M-M^\star)(1+\alpha)} \, \exp \left( -10^{-0.4(M-M^\star)}\right) \, ,
\end{equation}
where $M=\Muv$ is the far-UV absolute magnitude, $\alpha$ is the slope of the power law at low luminosities, $M^\star$ is the characteristic magnitude and $\phi^\star$ the normalisation. 

We also show in Fig.~\ref{fig:UV_LF_G} the far-UV luminosity function computed by \citet{Bouwens2014} using 598 galaxies at $z\sim7$ (black triangles)  and 225 galaxies at $z\sim8$ (black square). Fig.~\ref{fig:UV_LF_G} shows that our simple galaxy formation model (see Section~\ref{sec:Mutch}) reproduces well the observed far-UV luminosity function at $z=7$ and 8. This is quite a remarkable result, as by tuning the free parameters of the model to match the data at $z=7$ we naturally obtain prediction in agreement with $z=8$ observations. This indicates that our simple model, based on just three adjustable parameters, is able to subsume the complex physics of star formation at these redshifts, and that there is no need for a redshift evolution of the model parameters from $z=7$ to 8. We note that the \citet{Bouwens2014} data are not corrected for dust attenuation, but, as we discuss in Section~\ref{sec:dust}, we dot not expect dust corrections to be significant at these redshifts for most galaxy luminosities considered here. 

It is important to highlight that although we calibrate the galaxy formation model with the predictions from the Gaussian simulation, this calibration is valid for the non-Gaussian simulations too. The reason is that the effect of the initial non-Gaussianities adopted in this work decreases with decreasing redshift, hence the galaxy luminosity functions at $z = 7$ and 8 are almost indistinguishable among the different simulations (see Fig.~\ref{fig:UV_LF}). This guarantees the consistency of the non-Gaussian simulations with the \citet{Bouwens2014} observational constraints.
 
\begin{figure}
	\centering
	\resizebox{\hsize}{!}{\includegraphics{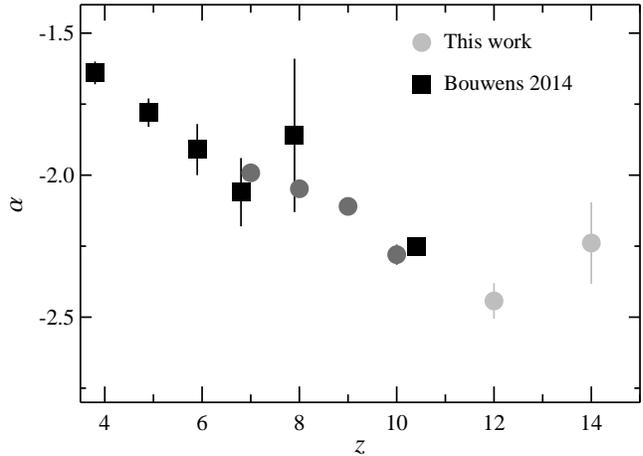}}
	\caption{Evolution with redshift of the faint-end slope of the far-UV luminosity function. Black squares indicate the slope of the UV luminosity function as measured by \citet{Bouwens2014} at $z=7$ and 8, while grey circles show the value of $\alpha$ computed from our galaxy formation model applied to the dark matter simulation with Gaussian initial conditions. In light grey, we indicate the allowed values of $\alpha$, which are more uncertain because of the combined effects of the low value of the characteristic far-UV magnitude and limited resolution of our simulation (see Section~\ref{sec:UV_LF}). Note that \citet{Bouwens2014} fix the value of $\alpha=-2.25$ at $z\sim10$ when fitting a Schechter function to their data.}
	\label{fig:alpha_z}
\end{figure}

\begin{figure*}
	\centering
	\resizebox{\hsize}{!}{\includegraphics{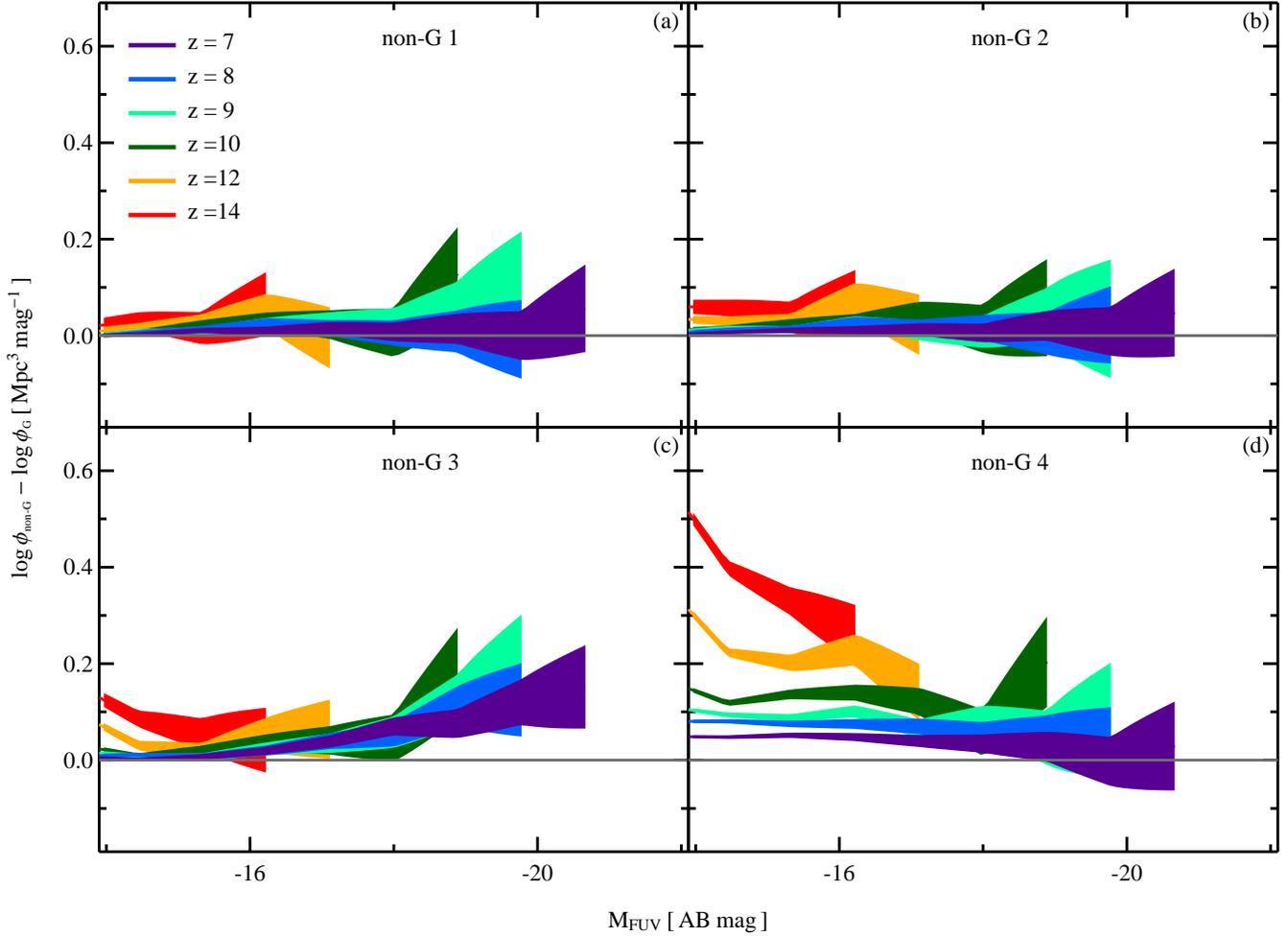}}
	\caption{Impact of primordial non-Gaussianities on the far-UV luminosity function. (a) Difference between the far-UV luminosity function computed from the non-Gaussian simulation 1 and that computed from the Gaussian simulation, at redshift $z=7$ (purple region), 8 (blue), 9 (light green), 10 (dark green), 12 (orange), 14 (red). (b) Same as (a), but for the non-Gaussian simulation 2. (c) Same as (a), but for the non-Gaussian simulation 3. (d) Same as (a), but for the non-Gaussian simulation 4. Each coloured region is computed assuming independent Poisson errors for each simulation and mass bin.}
	\label{fig:UV_LF}
\end{figure*}

Fig.~\ref{fig:UV_LF_G} shows that the number density of galaxies decreases with increasing redshift, at fixed far-UV absolute magnitude, in a qualitatively similar way to what we have already shown for the halo and stellar mass function (see Sec.~\ref{sec:haloMF}--\ref{sec:galMF}). The characteristic far-UV magnitude also decreases with increasing redshift, from $M_\txn{FUV}^\star=-21.3\pm0.37$ at $z=7$ to $M_\txn{FUV}^\star=-16.7\pm0.4$ at $z=14$, indicating that luminous galaxies become rarer and rarer at higher redshift. The faint-end end of the luminosity function steepens with increasing redshift, from $\alpha=-1.99\pm0.01$ at $z=7$ to $\alpha=-2.28\pm0.04$ at $z=10$ (see Fig.~\ref{fig:alpha_z}). At higher redshift, the low values of the characteristic magnitude make the exponential cut-off important at all stellar magnitudes considered here, therefore making the value of $\alpha$ in the Schechter function fit more uncertain. The flattening of the far-UV luminosity function from high to low redshift reflects the one already observed for the halo and stellar mass functions, and is likely caused by the same mechanism, i.e. galaxy merging.

A crucial factor for any model of Universe reionization based on star-forming galaxies is the faint-end slope of the far-UV luminosity function, since it determines the relative `weight' of low- and high-mass galaxies on the production rate of hydrogen-ionizing photons (see Equation~\ref{eq:rhoUV}). For this reason, we compare in Fig.~\ref{fig:alpha_z} the predicted evolution with redshift of the faint-end slope of the luminosity function of our model with the measurements of \citet{Bouwens2014}. Fig.~\ref{fig:alpha_z} shows that our model predictions agree remarkably well with the \citet{Bouwens2014} measurements at $z\sim7-10$. At $z\gtrsim10,$ our predictions for the evolution of $\alpha$ are influenced by the exponential cut-off of the Schechter function and are therefore less reliable. We also note that the choice of \citet{Bouwens2014} to fix the value of $\alpha=-2.25$ when fitting a Schechter function to their $z=10$ galaxy candidates is supported by our model predictions. As we noted above for the far-UV luminosity function, the convergence of the predictions obtained from the non-Gaussian simulations towards those of the Gaussian simulation at $z\lesssim8$ guarantees the evolution of $\alpha$ with redshift of the non-G models to be in agreement with the \citet{Bouwens2014} measurements.

As for the halo and stellar mass functions, we show in Fig.~\ref{fig:UV_LF} the effect of primordial non-Gaussianities on the far-UV galaxy luminosity function. To accomplish this, we compute the far-UV luminosity function for the Gaussian simulation and each non-Gaussian simulation in the same way as we do for the halo mass function (see Section~\ref{sec:haloMF}), adopting 10 bins in the range $-22 \le \Muv \le -14 $, and considering only bins with $\ge20$ objects. 
%We then compute, at each redshift, the difference between the luminosity density of galaxies as a function of the far-UV magnitude for the Gaussian simulation and each non-Gaussian simulation. The different colours in Fig.~\ref{fig:UV_LF} correspond to different redshifts, and the width of the coloured regions indicates the error in each bin, which we compute by summing in quadrature the relative errors in the luminosity density for the Gaussian and non-Gaussian simulation, which are in turn computed assuming a Poisson distribution. Note that this can underestimate the errors in bins with a small number of objects, i.e. for bright galaxies.

Fig.~\ref{fig:UV_LF}(a) shows that the far-UV luminosity function computed from the non-G~1 simulation and that computed from the Gaussian simulation are consistent at all redshifts and magnitudes, as we already noted for the halo and stellar mass functions. 

Fig.~\ref{fig:UV_LF}(b) indicates that the number density of faint galaxies ($\Muv \gtrsim -17$) is marginally larger in the non-G~2 simulation than in the Gaussian simulation. As we have already noted for the halo and galaxy stellar mass functions (see Figs.~\ref{fig:Halo_MF}(b) and \ref{fig:Gal_MF}(b)), this effect becomes stronger, at fixed far-UV absolute magnitude, with increasing redshift, reaching $\delta \phi \sim 0.07$ at $z=14$. 

As we consider models with stronger initial non-Gaussianities, the effect of such non-Gaussianities on the far-UV luminosity function becomes stronger, as already noted for the halo and stellar mass functions. Fig.~\ref{fig:UV_LF}(c) indeed shows significant differences between the non-G~3 and Gaussian far-UV luminosity function. The number density of galaxies is larger for the non-G~3 model than for the Gaussian one both at the faint and bright end of the luminosity function, and at all redshifts. At the bright end of the luminosity function the large Poisson errors do not allow us to identify trend with redshift. At $\Muv \gtrsim -15$ the effect of non-Gaussianities becomes stronger with increasing redshift, at fixed far-UV absolute magnitude, reaching  $\delta \phi \sim 0.15$ at $z=14$. 

As for the halo and stellar mass functions, Fig.~\ref{fig:UV_LF}(d) shows that the non-G~4 model, which has the strongest initial non-Gaussianities, produces the largest differences in the far-UV luminosity function. The number density of galaxies is larger in the non-G~4 model than in the Gaussian model at all redshift and magnitudes. We also recover the same trends with redshift and mass (here luminosity) already found for the halo and stellar mass functions, as the differences introduced by non-Gaussianities increases with increasing redshift, at fixed far-UV absolute magnitude, reaching  $\delta \phi \sim 0.5$ at $z=14$. At fixed redshift, the difference between the non-G~4 and Gaussian far-UV luminosity function increases with increasing absolute magnitude, i.e. with decreasing galaxy luminosity. 
 
In the next section, we present the reionization model we adopt to study how the differences in the far-UV luminosity function introduced by primordial non-Gaussianities might affect the reionization history of the Universe.   
 
\section{Reionization model}\label{sec:reioniz_model}

We can describe cosmic reionization with a differential equation accounting for the competing processes of hydrogen ionization by Lyman-continuum photons (with $E>13.6$ eV) and hydrogen recombination \citep[e.g.][]{Madau1999}
\begin{equation}\label{eq:QHII}
\frac{d\QHII}{dt} = \frac{\ndotion}{\nH } - \frac{\QHII}{\trec} \, ,
\end{equation}
where \QHII\ expresses the volume filling fraction of ionized hydrogen, \ndotion\ the comoving production rate of hydrogen ionizing photons, \nH\ the comoving average number density of hydrogen atoms, and \trec\ the average recombination time of hydrogen. Note that Equation~\ref{eq:QHII} does not account for collisional ionization and implicitly assumes that the ionization sources are widely separated, as it mixes mass-averaged (ionization fraction) and volume-averaged (recombination time) quantities (see the discussion in section 5 of \citealt{Finlator2012}).

The comoving average number density of hydrogen atoms \nH\ (units of $\mathrm{cm}^{-3}$) can be expressed as
\begin{equation*}
\nH = \frac{\Xp \, \Omegab \, \rhoc}{\mH} \, ,
\end{equation*}
where $\Xp=0.75$ indicates the primordial mass-fraction of hydrogen \citep[e.g.][]{Coc2013},
$\rhoc = 1.8787 \times 10^{-29} h^{-2}~\mathrm{g}~\mathrm{cm}^{3}$ the critical density of the Universe,
$\Omegab=0.052$ the fractional baryon density (assuming \Planck\ values for $\Omegam=0.307$ and a baryon fraction $\fb=0.17$), and $\mH=1.6735\times10^{-24}~\mathrm{g}$ the hydrogen mass.

\begin{figure}
	\centering
	\resizebox{\hsize}{!}{\includegraphics{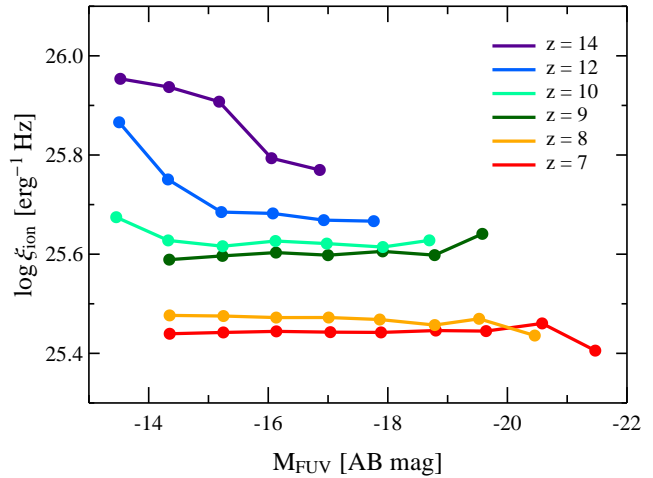}}
	\caption{Median value of the production rate of Lyman-continuum photons $\log \xiion$ as a function of the FUV absolute magnitude, at different redshifts, obtained from the spectral energy distributions of the galaxies in the Gaussian simulation. At fixed redshift, $\log \xiion$ exhibits variations $\lesssim 0.15$ dex in the luminosity range $-22 \le \Muv \le -14$.}
	\label{fig:xi_ion_z}
\end{figure}

We assume that reionization is driven solely by UV radiation emitted by massive stars in early galaxies, therefore neglecting any contribution from other ionization sources, such as Pop III stars and quasars (see Section~\ref{sec:agn_PopIII} for a discussion). Under this assumption, we can express the production rate of Lyman-continuum photons as
\begin{equation}\label{eqn:dniondt}
\ndotion = \fesc \, \xiion \, \rhoUV \, ,
\end{equation}
where \fesc\ is the fraction of Lyman-continuum photons escaping the interstellar medium (ISM) of the galaxy in which they are produced, \xiion\ the rate of Lyman-continuum photons per unit UV luminosity (computed at 1500 \AA), and \rhoUV\ the UV galaxy luminosity density. In the most general case, \fesc , \xiion\ and \rhoUV\ are redshift and luminosity dependent, but in this work we will consider \fesc\ and \xiion\ independent on the galaxy luminosity. This is motivated, for \xiion , by Fig.~\ref{fig:xi_ion_z}, which shows that the production rate of Lyman-continuum photons by galaxies with luminosity in the range $-22 \le \Muv \le -14$ varies by $\lesssim 0.15$ dex at fixed redshift. We can therefore adopt a constant value of \xiion\ in each redshift bin. In practice, we compute the median value of \xiion\ in each redshift bin for galaxies with luminosities in the range $-22 \le \Muv \le -14$, and report in Table~\ref{tab:xi_ion} the value obtained for the Gaussian and non-Gaussian simulations. As Table~\ref{tab:xi_ion} shows, at each redshift we obtain the same median value of \xiion\ for the Gaussian and non-Gaussian simulation 1, 2 and 3, and slightly lower values for the non-Gaussian simulation 4. 

\begin{table}
	\centering
	\begin{tabular}{c c c }
\toprule
\textbf{Redshift }	         & \multicolumn{2}{c}{$\boldsymbol{\log \xiion} \, [\txn{erg}^{-1}\txn{Hz}]$} \\

\cmidrule{2-3}

						         &  \textbf{G}, \textbf{non-G 1, 2 and 3} & \textbf{non-G 4} \\

\midrule

\vspace{\colhspace}
7	             &    25.44  &  25.43  \\
\vspace{\colhspace}
8				&      25.47 & 25.46  \\
\vspace{\colhspace}
9				&      25.59 & 25.57 \\
\vspace{\colhspace}
10				&      25.62 & 25.60 \\
\vspace{\colhspace}
12				&      25.72 & 25.68 \\
\vspace{\colhspace}
14				&      25.93 & 25.83 \\

\bottomrule
	\end{tabular}
	\caption{Production rate of hydrogen ionizing photons ($E>13.6$ eV) computed from the different Gaussian and non-Gaussian simulations. For each simulation, we compute the values listed in the table by dividing the galaxies in redshift bins, then considering the median value of  $\xiion$ for all galaxies with UV luminosity in the range  $-22 \le \Muv \le -14$.}
	\label{tab:xi_ion}	
\end{table}

The fraction of Lyman-continuum photons escaping their galaxies is one of the most uncertain ingredients of current reionization models, since there is no direct observational constraint on this quantity (see Section~\ref{sec:fesc} for a discussion). We therefore adopt two different scenarios, one in which the escape fraction is constant with redshift ($\fesc=0.2$), the other in which \fesc\ varies with redshift. We note that assuming an escape fraction increasing with redshift is equivalent to assuming \fesc\ increases with decreasing galaxy luminosity (or halo mass), once photoionization quenching and the evolution of the luminosity function are taken into account \citep{Alvarez2012}. We adopt the dependence of \fesc\ on redshift of \citet{Kuhlen2012} and \citet{Robertson2013}
\begin{equation}\label{eq:f_esc}
\fesc(z) = f_\mathrm{esc}^0 \left ( \frac{1+z}{5} \right )^k \, ,
\end{equation}
where we fix $f_\mathrm{esc}^0=0.054$ and $k=2.4$, following \citet{Robertson2013}. This equation implies an escape fraction $\fesc = 0.17$ at $z=7$ and $\fesc = 0.75$ at $z=14$, in qualitative agreement with the values obtained from state-of-the-art simulations by other groups \citep[e.g.][]{Paardekooper2013,Wise2014,So2014}.

We compute the UV luminosity density \rhoUV\ by analytically integrating the Schechter fit to the UV luminosity function obtained from our galaxy formation model applied to the different Gaussian and non-Gaussian simulation (see Section~\ref{sec:UV_LF}), obtaining
\begin{equation}\label{eq:rhoUV}
\begin{aligned}
\rhoUV & = \int_{-\infty}^{\MuvLim} \phi(M) L(M) dM \\
			& = \phi^\star \, M^\star \, \Gamma(\alpha+2, \MuvLim/ M^\star ) \, ,
\end{aligned}
\end{equation}
where $\Gamma(\alpha+2, \MuvLim/ \M^\star)$ is the upper Incomplete Gamma function. The result of the integration depends on the parameters of the Schechter function $\phi^\star$, $M^\star$ and $\alpha$, which we derive by means of an MCMC fitting of Equation~\ref{eq:Schechter_LF} to the binned luminosity function of each simulation at each redshift, and on the minimum galaxy luminosity  \MuvLim . 

The minimum galaxy luminosity is also an uncertain ingredient of any reionization model, as current observations of high redshift galaxies probe only the bright end of the UV luminosity function (see Section~\ref{sec:MuvLim} for a discussion). This value depends on the minimum mass of a halo able to cool down the gas to the temperatures required for star formation, and therefore on the highly uncertain physics of high redshift dwarf galaxy formation models (e.g. effects of cooling from molecular hydrogen and metal lines, UV background, stellar feedback, see \citealt{Wise2014}).  As for the escape fraction, we have to rely on simulations and previous work. We therefore fix $\MuvLim=-12$ (AB magnitude), which is, for instance, similar to the preferred value of \citet{Robertson2013}, and consistent with the recent simulations of \citet{Wise2014}, in which galaxies down to $\MuvLim \sim -5$ are formed. Since our model galaxies of $\Muv=-14$ correspond to haloes of $10^9\,\Msun$, adopting a limiting far-UV magnitude of -12 is equivalent to considering halo masses down to $\sim \txn 10^8\,\Msun$, assuming a fixed mass-to-light ratio in this mass range. Note that, unlike \citet{Robertson2013}, we do not adopt a single metallicity to compute the spectral energy distribution of our model galaxies, since we adopt the mass-metallicity relation for dwarf galaxies of \citet{Kirby2013} to calculate the chemical evolution of all galaxies in our simulations (see Section~\ref{sec:Mutch}).

The last quantity entering Equation~\ref{eq:QHII} is the average recombination time of ionized hydrogen atoms in the IGM, which is
\begin{equation}\label{eq:trec}
\begin{aligned}
\trec & = \lb \alphaB(T) \nE \rb^{-1} \\
		&	= \lb \CHII \alphaB(T) \fE \nH (1+z)^{3} \rb^{-1} \, , 																									
\end{aligned}
\end{equation}
where \alphaB\ is the hydrogen recombination coefficient, $\nE~=~\fE~\nH~(1+z)^3$ the number density of free electrons at redshift $z$, \fE\ the fraction of free electrons per hydrogen nucleus in the ionized IGM, and \CHII\ the `clumping factor', which accounts for inhomogeneities in the reionization process. We adopt an electron temperature of $T=20\,000$ K and adopt the case B recombination coefficient $\alphaB=2.52\times10^{-13}\,\txn{cm}^3\txn{s}^{-1}$.\footnote{Assuming an IGM temperature of $T=10\,000$ K would imply a case B recombination coefficient $\alphaB=2.59\times10^{-13}$, therefore a difference of $\lesssim3$ \% with respect to our choice.} Assuming that helium is doubly ionized at $z<4$ and singly ionized at higher redshift \citep[e.g.][]{Kuhlen2012}, we can express the number of free electrons per hydrogen nucleus as
\begin{equation*}
  %\label{eq:free_elec}
\fE=\left\{ \begin{array}{l l}
1 + \Yp/2\Xp & \hspace{1mm} \txn{at} \hspace{3mm} z \le 4 \,,\\
1 + \Yp/4\Xp & \hspace{1mm} \txn{at} \hspace{3mm} z > 4 \,,
\end{array}\right.
\label{eq_free_electr}
\end{equation*}
where \Xp\ and $\Yp=1-\Xp$ indicate the primordial mass fraction of hydrogen and helium, respectively. 

The clumping factor that enters the recombination time allows one to account for the effect of inhomogeneities in the density, temperature and ionization fields of the IGM. As for the escape fraction, one has to rely on simulations, since there are no observational constraints on the quantities determining the recombination rate of the IGM at high redshift (see Section~\ref{sec:CHII} for a discussion). We adopt for \CHII\ the analytic expression of \citet{Finlator2012}, which is based on state-of-the-art hydrodynamical simulations which include the effect of stellar feedback, photo-heating from a UV background, and self-shielding on the IGM:
\begin{equation}\label{eq:CHII}
  \CHII = 9.25 - 7.21\log{(1+z)} \, ,
\end{equation}
which implies a clumping factor that increases from 0.77 at $z=14$ to 3.16 at $z=6$. Note that Equation~\ref{eq:CHII} is in excellent agreement to that found by \citet{Wise2014} in their recent hydrodynamical simulations.

An important constraint on cosmic reionization comes from the optical depth of electrons to Thomson scattering, which can be expressed as
\begin{equation}\label{eq:tauE}
  \tauE = \int_0^\infty dz \, \frac{c\,(1+z)^2}{H(z)} \, \QHII(z) \, \sigma_T \, \nH \, \fE \, ,
\end{equation}  
where $c$ is the speed of light, $H(z)$ the Hubble parameter, \QHII\ the ionization fraction at redshift $z$,  $\sigma_T$ the cross-section of electrons to Thomson scattering, \nH\ the comoving average number density of hydrogen atoms and \fE\ the fraction of free electrons per hydrogen nucleus in the ionized IGM. 

\section{Implications for cosmic reionization}\label{sec:reioniz}

\begin{table}
	\centering
	\begin{tabular}{c c c}
\toprule

\textbf{Reionization model}	         & \textbf{$\boldsymbol{\fesc}$} & $\boldsymbol{\MuvLim}$ \\

\midrule

\vspace{\colhspace}
A	            				&   0.2  																						&  -12 		 \\
\vspace{\colhspace}
B							&      increasing with $z$ (see Equation~\ref{eq:f_esc}) 			& -12  	 \\
\vspace{\colhspace}
C							&      increasing with $z$ (see Equation~\ref{eq:f_esc}) 	 		& -7 	\\

\bottomrule
	\end{tabular}
	\caption{Different reionization models adopted in this work. We adopt for all models the clumping factor of \citet{Finlator2012} (see Equation~\ref{eq:CHII}), while we vary the escape fraction and limiting UV magnitude. Note that for model C we consider a limiting magnitude of -7, but adopting a constant number density of galaxies in the range $-12 \le \Muv \le -7$, equal to the density at $\Muv = -12$.}
	\label{tab:reioniz_models}	
\end{table}

We have shown in Section~\ref{sec:results} that introducing non-Gaussianities in the initial density perturbations which are then evolved by means of an N-body code produces measurable effects on the halo mass function, galaxy stellar mass function and far-UV luminosity function. These effects are marginally significant for the non-G~1 and non-G~2 models, while being stronger in the non-G~3 and non-G~4 simulations. This is caused by the different level of non-Gaussianities in the different models, as stronger deviations from purely Gaussian initial conditions produce stronger effects on the quantities  considered here. 
Two general features are shared by all non-Gaussian models considered in this work: the effect of primordial non-Gaussianities on the halo and stellar mass function and on the far-UV luminosity function increases with increasing redshift from $z=7$ to $z=14$; this effect becomes stronger at low halo and stellar masses, and thus in faint galaxies (except for the non-G~3 simulation, which show the same level of deviations at low and high masses). This is relevant to the Universe reionization history, since faint galaxies, i.e. galaxies which are currently unobservable because of their distance and low luminosity, are thought to be the primary source of ionizing radiation at high redshift. 

In this section, we will therefore explore the impact of different levels of primordial non-Gaussianities on the reionization history of the Universe and on the optical depth of electrons to Thomson scattering. This depends on the time-integral of the reionization history, and can be directly measured by CMB observations. We will also show how the emissivity rate of ionizing photons of our models is consistent with recent observations at $2\le z \le5$. 
To explore the effect of different assumptions about the reionization model on our results, we consider three different models, labelled `A', `B' and `C', in which we vary the escape fraction and limiting UV magnitude. Table~\ref{tab:reioniz_models} summarizes our choices for \fesc\ and \MuvLim\ for the three models: model A has $\fesc = 0.2$ at all redshifts and $\MuvLim=-12$; model B has the same \MuvLim\ as model A, but \fesc\ that increases with redshift following Equation~\ref{eq:f_esc}; model C has the same \fesc\ as model B, but a larger $\MuvLim=-7$, i.e. it includes fainter galaxies than model A and B. Note that to compute the UV luminosity density for model C (see Equation~\ref{eq:rhoUV}) we adopt a two-piece luminosity function: we consider a Schechter function till $\Muv=-12$, then a constant function in the range  $-12\le \Muv\le -7$. This is suggested by the recent hydro-dynamic simulation of \citet{Wise2014}, in which they show that the low efficiency of baryons conversion into stars in low-mass ($\log (\Mhalo/\Msun) \lesssim 8 $) haloes flattens the UV luminosity function at faint luminosities.

\subsection{Reionization history of the Universe}

We compute the  reionization history of the Universe for the different reionization models and both Gaussian and non-Gaussian simulations by numerically integrating Equation~\ref{eq:QHII}. For the fixed reionization model, the only quantity that varies in Equation~\ref{eq:QHII} among the Gaussian and non-Gaussian simulations is the production rate of ionizing photons, since it depends on \xiion , the rate of Lyman-continuum photons per unit UV luminosity, and on \rhoUV , the UV galaxy luminosity density. This in turns depends on the far-UV galaxy luminosity function through Equation~\ref{eq:rhoUV}, which we integrate analytically assuming for the parameters of the Schechter function $\phi^\star$, $M^\star$ and $\alpha$ the median of the posterior marginal distributions obtained by an MCMC fitting of the `numerical' luminosity function at each redshift (see Section~\ref{sec:UV_LF}). 

\begin{figure}
	\centering
	\resizebox{\hsize}{!}{\includegraphics{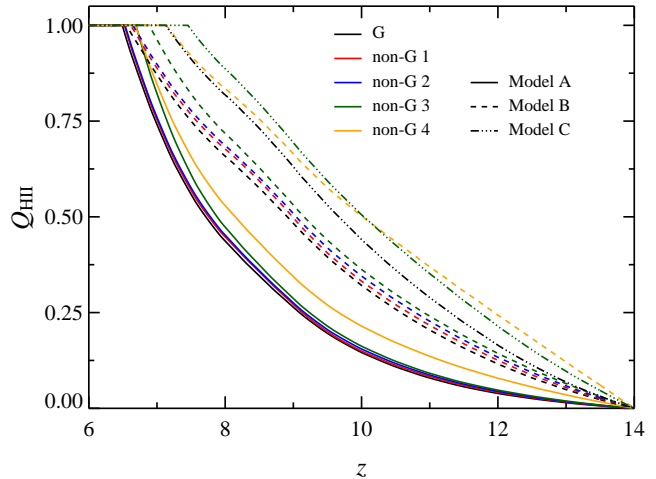}}
	\caption{Ionization fraction of the Universe as a function of redshift obtained by applying the different reionization models of Table~\ref{tab:reioniz_models} to the different Gaussian and non-Gaussian simulations. Different colours indicate different simulations: Gaussian (black), non-Gaussian 1 (red), non-Gaussian 2 (blue), non-Gaussian 3 (dark green) and non-Gaussian 4 (orange). Solid lines refer to the reionization model A ($\fesc=0.2$ and $\MuvLim=-12$), dashed lines to model B (\fesc\ increasing with $z$ as in Equation~\eqref{eq:f_esc} and $\MuvLim=-12$), dot-dashed lines to model C  (\fesc\ increasing with $z$ as in Equation~\eqref{eq:f_esc} and $\MuvLim=-7$). For clarity, for this latter model we just plot the results for the Gaussian and non-Gaussian 3 simulations.}
	\label{fig:reioniz_hist}
\end{figure}

We show in Fig.~\ref{fig:reioniz_hist} the results of the numerical integration of Equation~\ref{eq:QHII}, i.e. the fraction of ionized volume of the Universe as a function of redshift. Lines of different colours correspond to the different Gaussian and non-Gaussian simulations; solid lines refer to the reionization model A, dashed lines to model B, and dot-dashed lines to model C. 
 
The solid lines in Fig.~\ref{fig:reioniz_hist} indicate that the reionization histories obtained for the Gaussian, non-Gaussian 1, 2 and 3 simulations for the reionization model A show small differences with one another. This is a direct consequence of the similarity of the far-UV luminosity functions among the Gaussian, non-Gaussian 1, 2 and 3 simulations at redshift $z\lesssim12$, and therefore at most times (see Fig.~\ref{fig:UV_LF}). On the other hand, Fig.~\ref{fig:reioniz_hist} shows significant differences among the reionization history of the non-G~4 simulation and the other simulations. At a given redshift, the fraction of the IGM ionized  is larger for the non-G~4 simulation, being $\QHII = 0.08$ (0.53) at redshift $z=12$ (8), to be compared with $\QHII \sim 0.04$ (0.45) at the same redshifts for the other simulations. This difference is a direct consequence of the larger number of low-mass galaxies formed in the non-G~4 simulation with respect to the other simulations (see Fig.~\ref{fig:UV_LF}(d)), which, at each redshift, increases the number of photons available for hydrogen ionization.

The dashed lines in Fig.~\ref{fig:reioniz_hist} show the reionization histories for the different simulations obtained assuming the reionization model B, which, unlike model A, has an escape fraction that increases with increasing redshift. This makes the fraction of ionized IGM to increase more rapidly at high $z$, because of the larger number of photons available for hydrogen ionization at high redshift in model B than in model A. 
As for model A, the reionization histories of the Gaussian simulation and non-Gaussian simulation 1, 2 and 3 show smaller differences than that of the non-G~4 simulation. We note, however, that the dashed lines in Fig.~\ref{fig:reioniz_hist} are more separated than the solid ones, indicating that a model with an escape fraction that increases with redshift boosts the effect of primordial non-Gaussianities on the Universe reionization history. The reason is that in such a model the ionizing radiation emitted by higher redshift galaxies can escape the ISM more easily, hence increasing the contribution to Universe reionization of galaxies at high $z$, those most affected by primordial non-Gaussianities.

In Fig.~\ref{fig:reioniz_hist}, we show as dot-dashed lines \QHII\ for model C, which has a higher (fainter) \MuvLim\ than models A and B. For clarity, we just plot \QHII\ for the Gaussian and non-Gaussian 3 simulations. Fig.~\ref{fig:reioniz_hist} shows that considering fainter galaxies than those considered in models A and B makes \QHII\ increase faster at $z\gtrsim10$, while at same time increasing the difference between the Gaussian and non-Gaussian model 3. The explanation is similar to that given above for model B, but it appeals to the increasing effect of non-Gaussianities with decreasing galaxy luminosity, as adopting higher values of \MuvLim\ increases the weight of very faint galaxies, those most affected by non-Gaussianities, towards reionization.

\subsection{Electron Thomson scattering optical depth}
 
Direct measurements of the Universe ionized fraction through \Lya\ absorption from background quasars are effective up to neutral fractions $(1-\QHII)\sim10^{-3}$, since at higher values of $(1-\QHII)$ the resonance introduced by \Lya\ scattering makes ionizing photons almost completely absorbed by neutral hydrogen the IGM. This situation may change in the future through measurements of the 21 cm emission from the hyper-fine transition of neutral hydrogen, since this quantity directly depends on the hydrogen reionization history \citep[e.g. see the review of][]{Morales2010}.     
For now, one of the most important constraints on cosmic reionization comes from the optical depth of electrons to Thomson scattering $\tauE$, since this quantity depends on the (integrated) ionization fraction at different redshifts (see Equation~\eqref{eq:tauE}) and can be measured through CMB photons. 

\begin{figure}
	\centering
	\resizebox{\hsize}{!}{\includegraphics{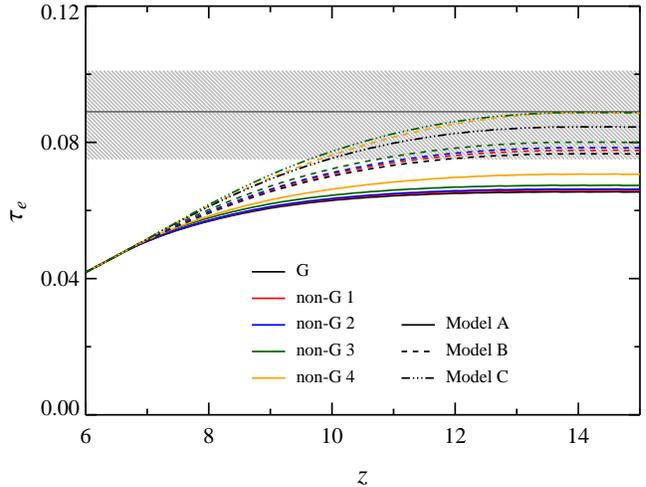}}
	\caption{Optical depth of electrons to Thomson scattering as a function of redshift obtained by numerically integrating Equation~\eqref{eq:tauE}. The dark-grey solid line indicates the `best-fit' value of \tauE\, along with the 68 \% confidence limits (grey hatched ragion), obtained by \citet{Planck2013_16} (see their Table 5, first column). Lines of different colours refer to different simulations: Gaussian (black), non-Gaussian 1 (red), non-Gaussian 2 (blue), non-Gaussian 3 (dark green) and non-Gaussian 4 (orange). Solid lines refer to the reionization model A ($\fesc=0.2$ and $\MuvLim=-12$), dashed lines to model B (\fesc\ increasing with $z$ as in Equation~\eqref{eq:f_esc} and $\MuvLim=-12$), dot-dashed lines to model C  (\fesc\ increasing with $z$ as in Equation~\eqref{eq:f_esc} and $\MuvLim=-7$). For clarity, for this latter model we just plot the results for the Gaussian and non-Gaussian 3 simulations.}
	\label{fig:electron_scatt}
\end{figure}

We show in Fig.~\ref{fig:electron_scatt} the optical depth of electrons to Thomson scattering for the different reionization models, for the Gaussian and non-Gaussian simulations, obtained by numerically integrating Equation~\ref{eq:tauE}. As in Fig.~\ref{fig:reioniz_hist}, different colours refer to different simulations, solid lines to the reionization model A, dashed lines to model B, and dot-dashed lines to model C. 

As for the reionization history shown in Fig.~\ref{fig:reioniz_hist}, solid lines in Fig.~\ref{fig:electron_scatt} indicate that the values of \tauE\ obtained from the Gaussian and non-Gaussian simulation 1, 2 and 3 show small differences when assuming a reionization model with constant escape fraction. As in Fig.~\ref{fig:reioniz_hist}, the non-G~4 simulation produces the largest difference in this quantity. This is not surprising, since the only term which varies in the computation of the electron optical depth is the fraction ionized at each redshift (see Equation~\eqref{eq:tauE}), which shows large variations between the non-G~4 simulation and the other simulations. Fig.~\ref{fig:reioniz_hist} shows also that assuming a constant escape fraction $\fesc=0.2$ produces an optical depth \tauE\ lower than the values currently allowed by \Planck\ observations, for all models. This suggests that a higher escape fraction and/or a fainter limiting UV magnitude are required to reionize earlier the Universe and obtain \tauE\ in agreement with \Planck\ constraints.

The dashed lines in Fig.~\ref{fig:electron_scatt} show \tauE\ for model B, in which the escape fraction increases with increasing redshift. Unlike model A, this one produces values of \tauE\ within the \Planck\ constraints for all Gaussian and non-Gaussian simulations.
As we have already highlighted for the reionization history, this model boosts the effect of primordial non-Gaussianities, increasing the differences in the Thomson scattering optical depth among all simulations. This is a direct consequence of the different reionization histories, and of the increased `weight' that an increasing \fesc\ with redshift gives to high redshift galaxies, those most affected by primordial non-Gaussianities.

The dot-dashed lines in Fig.~\ref{fig:electron_scatt} show the effect of increasing the limiting UV magnitude from $-12$ to $-7$ for the Gaussian and non-Gaussian~3 simulations. This increases, at a given redshift, the value of \tauE\ with respect to models A and B, while at the same time boosting the differences between the G and non-G~3 models. As we already noted for \QHII, the cause is the increased weight of very faint galaxies to reionization, that is of the galaxies most affected by non-Gaussianities.

\subsection{Ionizing emissivity}

We have shown in Fig.~\ref{fig:electron_scatt} that our models with variable escape fraction (models B and C) predict values of \tauE\ within \Planck\ constraints. However, this does not guarantee our models to be consistent with the (comoving) ionizing emissivity rate measured from the IGM opacity to \Lya\ photons. We therefore compute the ionizing emissivity rate of our reionization models, for the Gaussian and non-Gaussian simulations, and compare our results with the measures by \citet{Becker2013} of the ionizing emissivity at $2 \le z \le 5$. Note that in order to compare our predictions with the \citet{Becker2013} data, we have to extrapolate the predicted far-UV luminosity function to lower redshift, as our simulations stop at $z=6.5$. We therefore consider the relations between the free parameters of the Schechter function ($\phi^\star$, $M^\star$ and $\alpha$) and redshift, and note that these relations are linear in the range $6.5\le z \le9$. We therefore adopt a linear extrapolation to compute the far-UV luminosity function and ionizing emissivity rate down to $z=5$. 

\begin{figure}
	\centering
	\resizebox{\hsize}{!}{\includegraphics{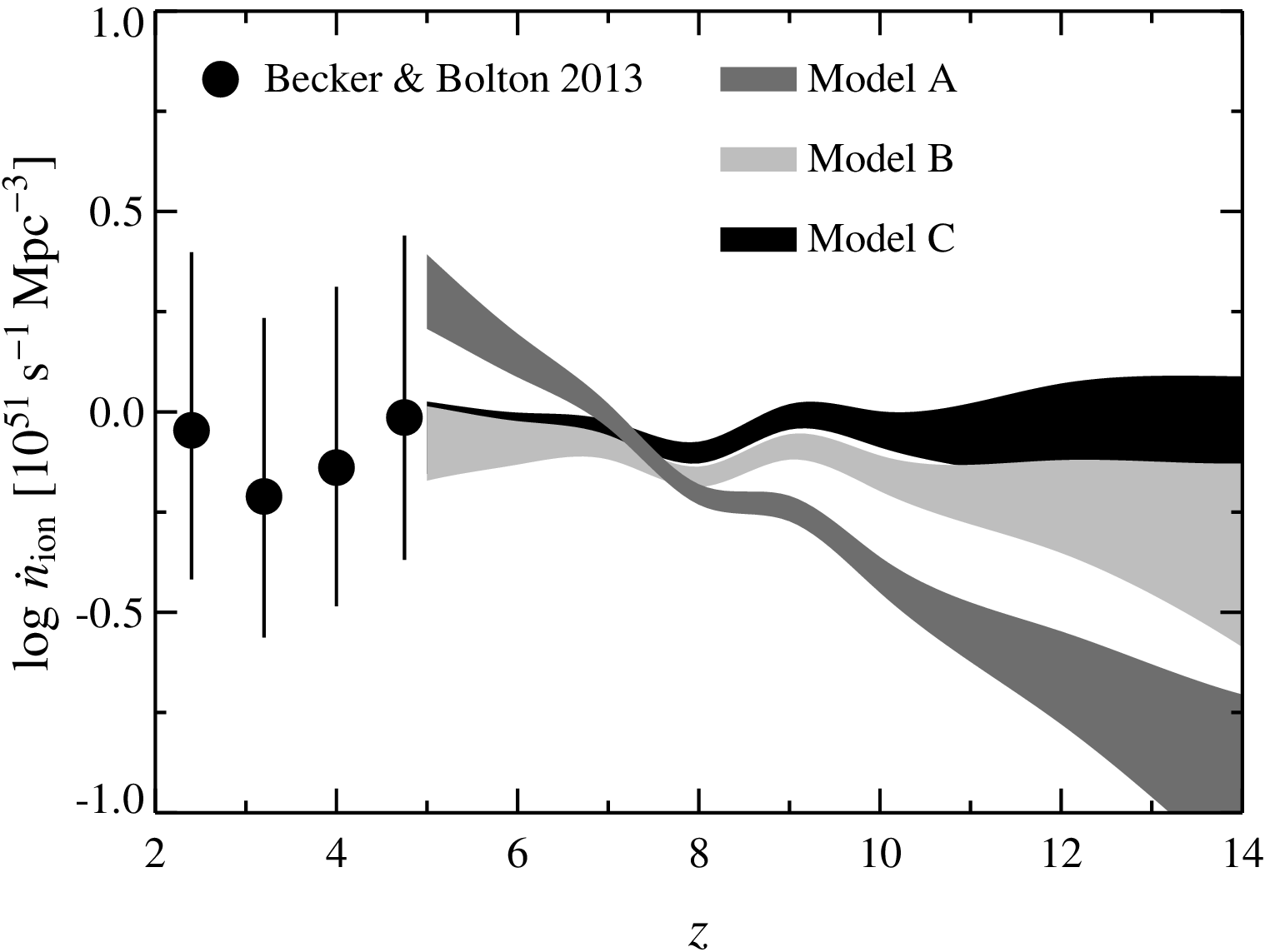}}
	\caption{Ionizing emissivity as a function of redshift for the different reionization models adopted in this work. Black circles indicate the ionizing emissivity measured by \citet{Becker2013} at $2\le z \le 5$. The shaded regions refer to different reionization models: model A (dark grey shaded region, $\fesc=0.2$ and $\MuvLim=-12$), model B (grey shaded region, \fesc\ increasing with $z$ as in Equation~\eqref{eq:f_esc} and $\MuvLim=-12$); model C (black shaded region, \fesc\ increasing with $z$ as in Equation~\eqref{eq:f_esc} and $\MuvLim=-7$). Note that the width of each shaded region reflects, for the fixed reionization model, the range spanned by the different Gaussian and non-Gaussian simulations, without accounting for other sources of errors (e.g. uncertainties in the luminosity function fit with a Schechter function, uncertainties in the free parameters of the reionization model). }
	\label{fig:emissivity}
\end{figure}

We show in Fig.~\ref{fig:emissivity} as shaded areas the emissivities predicted by the reionization models A (dark grey), B (grey), and C (black), while we indicate with black circles \citet{Becker2013} data. The width of the shaded areas at each redshift reflects the range of emissivities of the different Gaussian and non-Gaussian simulations. Fig.~\ref{fig:emissivity} shows that at high redshift our reionization models span a wide range of ionizing emissivities, while at $z=5$ the predictions of all models agree with \citet{Becker2013} observations. Fig.~\ref{fig:emissivity} also shows that the relative emissivities of models A, B and C are mainly driven by the adopted relation between escape fraction and redshift: at $z\gtrsim8$ the escape fraction is larger in models B and C than in model A, and so is the emissivity, while at $z\lesssim7$ the situation is reversed, as the escape fraction of model A becomes larger than that of models B and C.

\section{Uncertainties in the reionization model}\label{sec:uncert}

The model of cosmogical reionization that we have adopted depends on several assumptions, which we will discuss in this section. We firstly address our choice of neglecting the contribution of AGN and Pop III stars to cosmological reionization. Then, we discuss our assumptions about the adjustable parameters of the reionization model, namely the escape fraction, limiting UV magnitude and clumping factor. Finally, we justify our choice of not including dust attenuation in our model.

\subsection{Contribution of Active Galactic Nuclei and Pop III to hydrogen reionization}\label{sec:agn_PopIII}

In our model of cosmic reionization we have ignored the contribution of active galactic nuclei (AGN) and Pop III stars to the hydrogen ionizing budget; within our model the only sources of ionizing radiations are metal-enriched stars residing in galaxies. 

AGN do not contribute significantly to hydrogen reionization because their number density rapidly decreases at $z\gtrsim3$ \citep[e.g.][]{Hopkins2007,Ross2013}, and therefore they add few UV photons to those produced by massive stars \citep[e.g.][]{Faucher2009}. Nevertheless, \citet{Volonteri2009} have pointed out that secondary ionization from X-ray emission could boost AGN contribution to reionization up to 50-90 \% at $z\gtrsim8$. However, \citet{Grissom2014} have recently re-addressed this question, reaching a different conclusion. They consider the most recent constraints on the AGN X-ray luminosity function, and adopt a conservative model in which AGN accrete at the maximum (Eddington) rate and are unobscured. They consider primary and secondary ionization from X-ray photons, and find that the AGN contribution to Universe reionization is $\lesssim 14$ \% at all redshifts $6 \le z \le 14$ (see their figure 3).

Metal-free `Pop III' stars form from pristine gas which has not yet been enriched by previous generation of stars. The absence of metals and dust grains suggest that fragmentation of molecular clouds is less efficient, and simulations have indeed shown that the first stars form with an initial mass function richer in high mass stars than what observed in the Local Universe \citep[e.g.][]{McKee2008,Clark2011}, even though no observational constraints on their masses currently exist. Pop III stars are therefore thought to be massive and short-lived, and, despite their hot temperatures and hard-radiation spectrum, to give little contribution to the total budget of hydrogen ionizing photons (e.g. see figure 12 of \citealt{Wise2014} and figure 2 of \citealt{Paardekooper2013}). We note also that the chemical abundances of the IGM measured from damped \Lya\ systems at $z\lesssim6$ suggests a low contribution of Pop III stars to Universe reionization, as the yields of massive Pop III stars would leave peculiar imprints, which are not observed, on these abundances \citep[e.g.][]{Becker2012,Kulkarni2014}.

\subsection{Effect of changes in the adjustable parameters of the reionization model}

The reionization model we have presented in Section~\ref{sec:reioniz_model} depends on three adjustable parameters: the escape fraction of ionizing photons, limiting UV magnitude and clumping factor. These parameters are poorly constrained at high redshift by current observations, and therefore they are the major sources of uncertainty in modelling cosmological reionization. In this section, we discuss our choices for these three adjustable parameters, and how their variation will affect our results.

\subsubsection{Escape fraction}\label{sec:fesc}

The fraction of ionizing (Lyman-continuum) photons escaping the ISM of the galaxies in which they are produced alters the amount of photons available for hydrogen ionization, therefore affecting the reionization history of the Universe (see Equations~\ref{eq:QHII} and \ref{eqn:dniondt}). Observations can accurately constrain \fesc\ only in the local Universe, e.g. the (angle-averaged) Milky Way escape fraction is $\ave{\fesc} \sim 0.02$ \citep{Bland1999}. Adopting such a low value for the escape fraction would make Universe reionization by star-forming galaxies alone very challenging. At higher redshift constraints on \fesc\ are more uncertain, however data suggest that the escape fraction at $z\sim3$ is $\sim 0.05-0.07$ for Lyman-break galaxies (LBG), and $\sim 0.1-0.3$ for (fainter) Lyman-alpha emitters (LAE) (\citealt{Nestor2013}, see also \citealt{Jones2013}), pointing toward an increase of the escape fraction with increasing redshift.

Local constraints on \fesc\ are of limited use at high redshift, since the escape fraction depends, among other parameters, on galaxy morphologies, stellar feedback, and more in general on the interplay between galaxies and the IGM, while the available observational constraints at high redshift are still very uncertain. This has motivated in the last few years the development of extensive simulations to study how the escape fraction varies in dwarf galaxies at high redshift. 

Different groups have found consistent trends in their sophisticated hydrodynamical simulations of star-bursting dwarf galaxies \citep{Wise2009,Paardekooper2013,Wise2014,So2014,Kimm2014}. They find \fesc\ in these high redshift dwarf galaxies to be higher than in local disk galaxies ($\fesc > 0.1$ for high-$z$ dwarf), and to be tightly correlated to their star formation rate. Moreover, they find systematically larger escape fractions with decreasing halo mass. They explain these trends as follows: the irregular morphology of high-$z$ dwarf galaxies boosts \fesc\ with respect to local discs since star formation is not confined along the equatorial disc of the galaxies, hence UV radiation escaping star forming regions is less likely to be absorbed during its travel along the disc. Dwarf galaxies are also more affected by stellar feedback (i.e. ionization fronts, winds, shocks from supernovae) because of their shallower potential wells and irregular morphology. This also explains the trend they observe with halo mass, as feedback in lower mass haloes is more efficient in removing gas and creating cavities along which the ionizing radiation can escape. The dependence of \fesc\ on the star formation rate is also caused by stellar feedback, as a burst of star formation clears out the gas from the galaxy, allowing photons to escape more easily and increasing \fesc . 

To summarize, hydrodynamical simulations from different groups consistently show that the escape fraction is larger in high-$z$ dwarf galaxies than in low-$z$ discs; \fesc\ in high-$z$ galaxies is time- and space-dependent since it depends on the interplay between stellar feedback and gas in the ISM; \fesc\ increases with decreasing halo mass. Our choice of a constant escape fraction $\fesc=0.2$ at all halo masses and redshift is therefore conservative, while our model with \fesc\ increasing with redshift is based upon the above results, since, as the average halo mass decreases with increasing redshift, we expect the escape fraction to increase. This, as shown by the dashed lines in Figs.~\ref{fig:reioniz_hist}--\ref{fig:electron_scatt}, increases the effect of primordial non-Gaussianities on the Universe reionization history, since a larger fraction of the photons emitted by high redshift ($z\gtrsim 10$) galaxies, those most affected by primordial non-Gaussianities, becomes available for hydrogen ionization.

\subsubsection{Limiting UV magnitude}\label{sec:MuvLim}

The limiting UV magnitude affects the UV luminosity density (via the integral of Equation~\ref{eq:rhoUV}), and therefore the reionization history of the Universe (see Equations~\ref{eq:QHII} and \ref{eqn:dniondt}). This quantity depends on the minimum mass of a halo able to cool down the gas and sustain star formation. Constraining this quantity with data requires observing the faintest galaxies at high redshift, a task which will be very difficult even with the next generation of telescopes such as \textit{JWST} \citep[e.g. see fig 15 of][]{Wise2014}. This means that, as for the escape fraction, we have to rely on simulations to constrain the limiting UV magnitude. 

For a long time, it has been thought that only haloes hosting atomic hydrogen (i.e. with $\Tvir > 10^4$ K, or $\log(\Mvir/\Msun) \gtrsim 8$) can cool gas down to the temperatures required for stars to form \citep[e.g.][]{Bromm2003}. The reason is that earlier models  overpredicted the efficiency of  UV background (Lyman-Werner) radiation in dissociating hydrogen molecules, hence preventing them from being an effective gas coolant. In the past few years, however, several groups \citep[e.g.][]{Wise2007,Oshea2008} have shown by means of sophisticated hydrodynamical simulations that hydrogen molecules can survive even in the presence of an extreme UV background, and that they can therefore act as an effective coolant in haloes with masses $\log(\Mvir/\Msun) \lesssim 8$. In particular, \citet{Wise2014} show that haloes with masses as low as $\log (\Mhalo/\Msun) \sim 6.5$, corresponding to $\Muv \sim 5.5$, can host dense molecular gas able to form stars. The same authors find however an almost constant number density of faint galaxies ($\Muv \gtrsim -12$), unlike what one would expect by extrapolating the galaxy luminosity function to very faint luminosities. 

In our model, we have chosen a limiting UV magnitude $\MuvLim = -12$, similar to what has been assumed in previous works \citep[e.g.][]{Robertson2013} and in agreement with the \citet{Wise2014} predictions for an extrapolation of the luminosity function at faint magnitudes. We have also explored that the effect of increasing \MuvLim\ on the Universe reionization history: assuming $\MuvLim > -12$ increases the `weight' of low-mass galaxies to the reionization budget, thus boosting the effect of non-Gaussianities in a qualitatively similar way to what we observed for the escape fraction (see Fig.~\ref{fig:electron_scatt}). On the other hand, considering $\MuvLim < -12$ would reduce the effect primordial non-Gaussianities on cosmological reionization, although in this case, other sources of ionizing radiation would be required to match \Planck\ constraints on \tauE.

\subsubsection{Clumping factor}\label{sec:CHII}

The clumping factor accounts for inhomogeneities in the density, temperature and ionization fields of the IGM. These inhomogeneities make the recombination rate computed assuming a density and temperature averaged over the whole Universe different from that computed from averaging over HII regions, since this rate depends on the local temperature and density of electrons and protons. A clumping factor $\CHII > 1$ reduces the recombination time (see Equation~\ref{eq:trec}), delaying reionization with respect to a case in which $\CHII \le 1$. As for the escape fraction and limiting UV magnitude, the clumping factor too can only be studied only through simulations, since there are no observations of this quantity at the redshifts of interest for reionization.

A common definition of the clumping factor in simulations involves only the gas density $\CHII = \langle \rho^2 \rangle_\txn{IGM} / \langle \rho \rangle ^ 2$, where $\rho$ indicates the baryon density and brackets refer to a volume-average over the recombining IGM and over the whole Universe. By adopting such a definition, early simulations obtained large clumping factors $\CHII \sim 30$ \citep[e.g.][]{Gnedin1997}, which would make reionization hard to achieve with star-forming galaxies alone. 
More recent works consider a density threshold in order to separate gas in the ISM from gas in the IGM, accounting also for photo-ionization by a UV background. This lowers the clumping factor by almost an order of magnitude ($\CHII \sim 3$ at $z=6$), and predicts a decreasing \CHII\ with increasing redshift \citep{Pawlik2009}. However, a limitation of the model of \citet{Pawlik2009} is that it assumes an optically thin IGM, therefore not accounting for self-shielding in over-dense regions. 

This limitation has been overcome in the recent simulation of \citet{Finlator2012}, in which they include a sub-grid model for the self-shielding of over-dense regions in the IGM. They find that the density threshold for self-shielding evolves with redshift, as it depends on both the UV background and gas temperature, which in turns varies with $z$. Their main result is that accounting for both self-shielding and fluctuations in the IGM temperature further reduces the clumping factor with respect to \citet{Pawlik2009} values (see figure 3 of \citealt{Finlator2012}). We therefore adopt equation~8 of \citet{Finlator2012}, which is valid in the redshift range $5\le z \le 15$, to describe the evolution with redshift of the clumping factor in our reionization model.
This equation predicts a clumping factor which decreases with increasing redshift (see figure 11 of \citealt{Wise2014} for a similar prediction), i.e. a larger recombination time at high redshift with respect to a model in which \CHII\ is constant with redshift \citep[e.g. as in][]{Robertson2013}. Similarly to the effect of an escape fraction that increases with redshift, this boosts the effect of primordial non-Gaussianities on cosmological  reionization, since a lower clumping factor at high redshift increases the recombination time (see Equations~\ref{eq:QHII} and \ref{eq:trec}), hence enhancing the efficiency of high ($z\gtrsim10$) galaxies, those most affected by non-Gaussianities, in ionizing hydrogen.

\subsection{Dust attenuation}\label{sec:dust}

Simulations have shown that dust in high-$z$ galaxies has little effect on \fesc , as the absorption cross section of Lyman-continuum photons to dust is much smaller than to neutral hydrogen \citep[e.g.][]{Gnedin2008,Paardekooper2011}. We have also neglected dust attenuation when matching our predictions for the far-UV luminosity function with the \citet{Bouwens2014} observations at $z=7$ and 8. This is supported by the recent work of \citet{Wilkins2013}, who consider the UV continuum slope of a large sample of star-forming galaxies at $z\gtrsim5$. They estimate the amount of UV attenuation affecting their sample by considering different attenuation and extinction curves, so as to minimise the model-dependence of their results. They find that UV attenuation decreases with increasing redshift, at fixed UV luminosity, so that at $z\sim6$ the mean far-UV attenuation of galaxies with $\Muv\sim-21.5$ is in the range $0.5-1.5$, decreasing to $0.5-1$ at $z=7$, with the exact values depending on the adopted curve. They also find that \Auv\ decreases with decreasing UV luminosity, at fixed redshift, so that at $z=7$ the far-UV attenuation is in the range $0-0.5$ for galaxies with $\Muv \lesssim -20.3$. This suggests that dust corrections are not important for most galaxy luminosities and redshifts considered in this work.

\section{Conclusions}\label{sec:conclus}

Understanding the details of cosmological  reionization is one of the big challenges of current high redshift astronomy. Observations have established that at high redshift ($z \gtrsim 6$), low-mass galaxies outnumbered high-mass galaxies more than at lower redshift, i.e. that the UV luminosity function becomes steeper with increasing redshift \citep[e.g.][]{Bouwens2014}. This, combined with the observed decline of the AGN number density at $z\gtrsim3$ and the predicted negligible contribution to the ionizing budget of Pop III stars, suggests that dwarf galaxies at $z \gtrsim 6$ are likely responsible for cosmological  reionization. 

%The reason is the observed steepening of the faint-end of the far-UV galaxy luminosity function with increasing redshift, which implies that at high redshift ($z \gtrsim 6$) low-mass galaxies outnumbered high-mass ones more than at lower redshift \citep[e.g.][]{Bouwens2014}. Also, state-of-the-art hydrodynamical simulations of high redshift galaxies have consistently shown that hydrogen ionizing photons can escape more easily from low-mass than high-mass haloes \citep[e.g.][]{Wise2009,Paardekooper2013,Wise2014,So2014,Kimm2014}. This makes the reionization history of the Universe depend on the number density of dwarf galaxies at high redshift. 

In this work, we have extended the analysis of paper~I on the impact of scale-dependent non-Gaussianities by considering their effects on the UV galaxy luminosity function and on the reionization history of the Universe. Specifically, we have run 5 cosmological N-body simulations with identical power spectra, but different scale-dependent primordial non-Gaussianities, all consistent with \Planck\ constraints. As in paper~I, we have shown that models with stronger initial non-Gaussianities produce larger effects (up to a factor of 3) on the halo mass function, and that this effect increases with increasing redshift, at fixed halo mass. We find in all simulations except the one with the steepest scale-dependent non-Gaussianity that the effect of such primordial non-Gaussianities is more significant in low- than high-mass haloes, at fixed redshift. % , although this result is influenced by the different Poisson noise affecting the low and high mass end of the mass function.

%As the production rate of hydrogen ionizing photons depends on the far-UV galaxy luminosity function, and in particular on its faint-end slope, we have calibrated our galaxy formation model to reproduce the most recent determination of the far-UV luminosity function at $z=7$ and 8 \citep{Bouwens2014}. 

By combining the halo merger trees obtained from simulations with different initial conditions  together with a modified version of the galaxy formation model of \citet{Mutch2013} and the \citet{BC03} population synthesis code, we have then matched the most recent determination of the far-UV galaxy luminosity function at $z=7$ and 8. We have also shown that the prediction of our model for the evolution of the faint-end slope of the far-UV galaxy luminosity function is in excellent agreement with observations at $z=6-10$. This demonstrates that a phenomenological galaxy formation model which subsumes the complex baryonic physics in a few analytic functions is accurate enough to predict the observed properties of the galaxy population at high redshift.  

Having calibrated in such a way the galaxy formation model, we apply it to the different simulations with non-Gaussian initial conditions, showing that the effects introduced by non-Gaussianities on the distribution of halo masses propagate to the galaxy stellar mass function and far-UV luminosity function. In particular, the initial non-Gaussianities considered in this work increase the number density of faint galaxies (up to a factor of 3), and this effect becomes stronger with increasing redshift, at fixed UV magnitude and similarly to what noted for the halo mass function. 

Finally, we have appealed to an analytic reionization model to quantify the effect of primordial non-Gaussianities on the reionization history of the Universe. We find that the effect of such non-Gaussianities depends on the adopted reionization model, and in particular on the adopted values for the ionizing escape fraction and limiting UV magnitude. For a given set of parameters describing the far-UV luminosity function, these two quantities, along with the clumping factor, determine the relative contribution to Universe reionization of galaxies at different redshifts, and hence the impact of primordial non-Gaussianities, as their effect is strongly redshift-dependent. 

We find that adopting a (redshift-dependent) ionizing escape fraction and clumping factor, as predicted by state-of-the-art hydrodynamical simulations, boosts the imprint of primordial non-Gaussianities on the reionization history of the Universe and on the electron Thomson scattering optical depth. The same qualitative effect is produced by decreasing the limiting UV luminosity, i.e. by considering stars residing in `mini-haloes' (down to $\log(\Mhalo/\Msun)\sim6$, as in \citealt{Wise2014}). We also find that our reionization models with ionizing escape fraction increasing with redshift produce \tauE\ in agreement with \Planck\ constraints, while at the same time allowing us to match the ionizing emissivity rate measured at $z\sim5$. Although current uncertainties on the physics of reionization and on the determination of \tauE\ dominate the signal of non-Gaussianities, more accurate measurements of this quantity, combined to a better understanding of the ingredients of reionization models, have the potential to constrain the shape of primordial density fluctuations, and may eventually be used to narrow down the allowed model-space of inflation.

\section*{Acknowledgments}

We thank the anonymous referee for insightful comments which helped improving the paper. We are grateful to Kristian Finlator and Stephane Charlot for their comments on an early draft of this paper, which helped us improving the work. JC acknowledges the support of the European Research Council via  Advanced Grant $\txn{N}^\txn{o}$ 267117 (DARK, P.I. Joseph Silk) and via Advanced Grant $\txn{N}^\txn{o}$ 321323 (NEOGAL, P.I. Stephane Charlot). TN is supported by the Japan Society for the Promotion of Science (JSPS) Postdoctoral Fellowships for Research Abroad.

\bibliographystyle{mn2e} % style aa.bst
\bibliography{Bib_nonG_LF} % your references Yourfile.bib

\begin{thebibliography}{78}
\expandafter\ifx\csname natexlab\endcsname\relax\def\natexlab#1{#1}\fi

\bibitem[{{Alishahiha} {et~al}\mbox{.}(2004){Alishahiha}, {Silverstein}, \&
  {Tong}}]{Alishahiha2004}
{Alishahiha} M., {Silverstein} E., {Tong} D., 2004, \prd, 70, 123505

\bibitem[{{Alvarez} {et~al}\mbox{.}(2012){Alvarez}, {Finlator}, \&
  {Trenti}}]{Alvarez2012}
{Alvarez} M.~A., {Finlator} K., {Trenti} M., 2012, \apjl, 759, L38

\bibitem[{{Aubert} {et~al}\mbox{.}(2004){Aubert}, {Pichon}, \&
  {Colombi}}]{Aubert2004}
{Aubert} D., {Pichon} C., {Colombi} S., 2004, \mnras, 352, 376

\bibitem[{{Becker} {et~al}\mbox{.}(2011){Becker}, {Huterer}, \&
  {Kadota}}]{Becker2011}
{Becker} A., {Huterer} D., {Kadota} K., 2011, \jcap, 1, 6

\bibitem[{{Becker} \& {Bolton}(2013)}]{Becker2013}
{Becker} G.~D., {Bolton} J.~S., 2013, \mnras, 436, 1023

\bibitem[{{Becker} {et~al}\mbox{.}(2012){Becker}, {Sargent}, {Rauch}, \&
  {Carswell}}]{Becker2012}
{Becker} G.~D., {Sargent} W.~L.~W., {Rauch} M., {Carswell} R.~F., 2012, \apj,
  744, 91

\bibitem[{{Behroozi} {et~al}\mbox{.}(2013){Behroozi}, {Wechsler}, \&
  {Conroy}}]{Behroozi2013}
{Behroozi} P.~S., {Wechsler} R.~H., {Conroy} C., 2013, \apj, 770, 57

\bibitem[{{Berlind} \& {Weinberg}(2002)}]{Berlind2002}
{Berlind} A.~A., {Weinberg} D.~H., 2002, \apj, 575, 587

\bibitem[{{Berlind} {et~al}\mbox{.}(2003){Berlind}, {Weinberg}, {Benson},
  {Baugh}, {Cole}, {Dav{\'e}}, {Frenk}, {Jenkins}, {Katz}, \&
  {Lacey}}]{Berlind2003}
{Berlind} A.~A. {et~al.}, 2003, \apj, 593, 1

\bibitem[{{Bland-Hawthorn} \& {Maloney}(1999)}]{Bland1999}
{Bland-Hawthorn} J., {Maloney} P.~R., 1999, \apjl, 510, L33

\bibitem[{{Bouwens} {et~al}\mbox{.}(2014){Bouwens}, {Illingworth}, {Oesch},
  {Trenti}, {Labbe'}, {Bradley}, {Carollo}, {van Dokkum}, {Gonzalez},
  {Holwerda}, {Franx}, {Spitler}, {Smit}, \& {Magee}}]{Bouwens2014}
{Bouwens} R.~J. {et~al.}, 2014, ArXiv e-prints

\bibitem[{{Bromm} \& {Loeb}(2003)}]{Bromm2003}
{Bromm} V., {Loeb} A., 2003, \apj, 596, 34

\bibitem[{{Bruzual} \& {Charlot}(2003)}]{BC03}
{Bruzual} G., {Charlot} S., 2003, \mnras, 344, 1000

\bibitem[{{Cen} \& {Ostriker}(2006)}]{Cen2006}
{Cen} R., {Ostriker} J.~P., 2006, \apj, 650, 560

\bibitem[{{Chabrier}(2003)}]{Chabrier2003}
{Chabrier} G., 2003, \pasp, 115, 763

\bibitem[{{Chen}(2005)}]{Chen2005}
{Chen} X., 2005, \prd, 72, 123518

\bibitem[{{Clark} {et~al}\mbox{.}(2011){Clark}, {Glover}, {Klessen}, \&
  {Bromm}}]{Clark2011}
{Clark} P.~C., {Glover} S.~C.~O., {Klessen} R.~S., {Bromm} V., 2011, \apj, 727,
  110

\bibitem[{{Coc} {et~al}\mbox{.}(2013){Coc}, {Uzan}, \& {Vangioni}}]{Coc2013}
{Coc} A., {Uzan} J.-P., {Vangioni} E., 2013, ArXiv e-prints

\bibitem[{{Cole} {et~al}\mbox{.}(2000){Cole}, {Lacey}, {Baugh}, \&
  {Frenk}}]{Cole2000}
{Cole} S., {Lacey} C.~G., {Baugh} C.~M., {Frenk} C.~S., 2000, \mnras, 319, 168

\bibitem[{{Conroy} {et~al}\mbox{.}(2006){Conroy}, {Wechsler}, \&
  {Kravtsov}}]{Conroy2006}
{Conroy} C., {Wechsler} R.~H., {Kravtsov} A.~V., 2006, \apj, 647, 201

\bibitem[{{Crocce} {et~al}\mbox{.}(2006){Crocce}, {Pueblas}, \&
  {Scoccimarro}}]{Crocce2006}
{Crocce} M., {Pueblas} S., {Scoccimarro} R., 2006, \mnras, 373, 369

\bibitem[{{Crociani} {et~al}\mbox{.}(2009){Crociani}, {Moscardini}, {Viel}, \&
  {Matarrese}}]{Crociani2009}
{Crociani} D., {Moscardini} L., {Viel} M., {Matarrese} S., 2009, \mnras, 394,
  133

\bibitem[{{Croton} {et~al}\mbox{.}(2006){Croton}, {Springel}, {White}, {De
  Lucia}, {Frenk}, {Gao}, {Jenkins}, {Kauffmann}, {Navarro}, \&
  {Yoshida}}]{Croton2006}
{Croton} D.~J. {et~al.}, 2006, \mnras, 365, 11

\bibitem[{{Fan} {et~al}\mbox{.}(2006){Fan}, {Carilli}, \& {Keating}}]{Fan2006}
{Fan} X., {Carilli} C.~L., {Keating} B., 2006, \araa, 44, 415

\bibitem[{{Faucher-Gigu{\`e}re} {et~al}\mbox{.}(2009){Faucher-Gigu{\`e}re},
  {Lidz}, {Zaldarriaga}, \& {Hernquist}}]{Faucher2009}
{Faucher-Gigu{\`e}re} C.-A., {Lidz} A., {Zaldarriaga} M., {Hernquist} L., 2009,
  \apj, 703, 1416

\bibitem[{{Finlator} {et~al}\mbox{.}(2012){Finlator}, {Oh}, {{\"O}zel}, \&
  {Dav{\'e}}}]{Finlator2012}
{Finlator} K., {Oh} S.~P., {{\"O}zel} F., {Dav{\'e}} R., 2012, \mnras, 427,
  2464

\bibitem[{{Fukugita} {et~al}\mbox{.}(1998){Fukugita}, {Hogan}, \&
  {Peebles}}]{Fukugita1998}
{Fukugita} M., {Hogan} C.~J., {Peebles} P.~J.~E., 1998, \apj, 503, 518

\bibitem[{{Gnedin} {et~al}\mbox{.}(2008){Gnedin}, {Kravtsov}, \&
  {Chen}}]{Gnedin2008}
{Gnedin} N.~Y., {Kravtsov} A.~V., {Chen} H.-W., 2008, \apj, 672, 765

\bibitem[{{Gnedin} \& {Ostriker}(1997)}]{Gnedin1997}
{Gnedin} N.~Y., {Ostriker} J.~P., 1997, \apj, 486, 581

\bibitem[{{Grissom} {et~al}\mbox{.}(2014){Grissom}, {Ballantyne}, \&
  {Wise}}]{Grissom2014}
{Grissom} R.~L., {Ballantyne} D.~R., {Wise} J.~H., 2014, \aap, 561, A90

\bibitem[{{Habouzit} {et~al}\mbox{.}(2014){Habouzit}, {Nishimichi}, {Peirani},
  {Mamon}, {Silk}, \& {Chevallard}}]{Habouzit2014}
{Habouzit} M., {Nishimichi} T., {Peirani} S., {Mamon} G.~A., {Silk} J.,
  {Chevallard} J., 2014, ArXiv e-prints

\bibitem[{{Henriques} {et~al}\mbox{.}(2013){Henriques}, {White}, {Thomas},
  {Angulo}, {Guo}, {Lemson}, \& {Springel}}]{Henriques2013}
{Henriques} B.~M.~B., {White} S.~D.~M., {Thomas} P.~A., {Angulo} R.~E., {Guo}
  Q., {Lemson} G., {Springel} V., 2013, \mnras, 431, 3373

\bibitem[{{Hopkins} {et~al}\mbox{.}(2007){Hopkins}, {Richards}, \&
  {Hernquist}}]{Hopkins2007}
{Hopkins} P.~F., {Richards} G.~T., {Hernquist} L., 2007, \apj, 654, 731

\bibitem[{{Jones} {et~al}\mbox{.}(2013){Jones}, {Ellis}, {Schenker}, \&
  {Stark}}]{Jones2013}
{Jones} T.~A., {Ellis} R.~S., {Schenker} M.~A., {Stark} D.~P., 2013, \apj, 779,
  52

\bibitem[{{Kauffmann} {et~al}\mbox{.}(1993){Kauffmann}, {White}, \&
  {Guiderdoni}}]{Kauffmann1993}
{Kauffmann} G., {White} S.~D.~M., {Guiderdoni} B., 1993, \mnras, 264, 201

\bibitem[{{Kimm} \& {Cen}(2014)}]{Kimm2014}
{Kimm} T., {Cen} R., 2014, \apj, 788, 121

\bibitem[{{Kirby} {et~al}\mbox{.}(2013){Kirby}, {Cohen}, {Guhathakurta},
  {Cheng}, {Bullock}, \& {Gallazzi}}]{Kirby2013}
{Kirby} E.~N., {Cohen} J.~G., {Guhathakurta} P., {Cheng} L., {Bullock} J.~S.,
  {Gallazzi} A., 2013, \apj, 779, 102

\bibitem[{Komatsu \& Spergel(2001)}]{Komatsu2001}
Komatsu E., Spergel D.~N., 2001, Phys.Rev., D63, 063002

\bibitem[{{Kuhlen} \& {Faucher-Gigu{\`e}re}(2012)}]{Kuhlen2012}
{Kuhlen} M., {Faucher-Gigu{\`e}re} C.-A., 2012, \mnras, 423, 862

\bibitem[{{Kulkarni} {et~al}\mbox{.}(2014){Kulkarni}, {Hennawi}, {Rollinde}, \&
  {Vangioni}}]{Kulkarni2014}
{Kulkarni} G., {Hennawi} J.~F., {Rollinde} E., {Vangioni} E., 2014, \apj, 787,
  64

\bibitem[{{Lo Verde} {et~al}\mbox{.}(2008){Lo Verde}, {Miller}, {Shandera}, \&
  {Verde}}]{LoVerde2008}
{Lo Verde} M., {Miller} A., {Shandera} S., {Verde} L., 2008, JCAP, 4, 14

\bibitem[{{Lu} {et~al}\mbox{.}(2011){Lu}, {Mo}, {Weinberg}, \& {Katz}}]{Lu2011}
{Lu} Y., {Mo} H.~J., {Weinberg} M.~D., {Katz} N., 2011, \mnras, 416, 1949

\bibitem[{{Madau} {et~al}\mbox{.}(1999){Madau}, {Haardt}, \&
  {Rees}}]{Madau1999}
{Madau} P., {Haardt} F., {Rees} M.~J., 1999, \apj, 514, 648

\bibitem[{{Matarrese} {et~al}\mbox{.}(2000){Matarrese}, {Verde}, \&
  {Jimenez}}]{Matarrese2000}
{Matarrese} S., {Verde} L., {Jimenez} R., 2000, \apj, 541, 10

\bibitem[{{McKee} \& {Tan}(2008)}]{McKee2008}
{McKee} C.~F., {Tan} J.~C., 2008, \apj, 681, 771

\bibitem[{{Morales} \& {Wyithe}(2010)}]{Morales2010}
{Morales} M.~F., {Wyithe} J.~S.~B., 2010, \araa, 48, 127

\bibitem[{{Moster} {et~al}\mbox{.}(2013){Moster}, {Naab}, \&
  {White}}]{Moster2013}
{Moster} B.~P., {Naab} T., {White} S.~D.~M., 2013, \mnras, 428, 3121

\bibitem[{{Mutch} {et~al}\mbox{.}(2013){Mutch}, {Croton}, \&
  {Poole}}]{Mutch2013}
{Mutch} S.~J., {Croton} D.~J., {Poole} G.~B., 2013, \mnras, 435, 2445

\bibitem[{{Nestor} {et~al}\mbox{.}(2013){Nestor}, {Shapley}, {Kornei},
  {Steidel}, \& {Siana}}]{Nestor2013}
{Nestor} D.~B., {Shapley} A.~E., {Kornei} K.~A., {Steidel} C.~C., {Siana} B.,
  2013, \apj, 765, 47

\bibitem[{{Nishimichi}(2012)}]{Nishimichi2012}
{Nishimichi} T., 2012, JCAP, 8, 37

\bibitem[{{Nishimichi} {et~al}\mbox{.}(2009){Nishimichi}, {Shirata}, {Taruya},
  {Yahata}, {Saito}, {Suto}, {Takahashi}, {Yoshida}, {Matsubara}, {Sugiyama},
  {Kayo}, {Jing}, \& {Yoshikawa}}]{Nishimichi2009}
{Nishimichi} T. {et~al.}, 2009, \pasj, 61, 321

\bibitem[{{Nishimichi} {et~al}\mbox{.}(2010){Nishimichi}, {Taruya}, {Koyama},
  \& {Sabiu}}]{Nishimichi2010}
{Nishimichi} T., {Taruya} A., {Koyama} K., {Sabiu} C., 2010, \jcap, 7, 2

\bibitem[{{Oesch} {et~al}\mbox{.}(2013){Oesch}, {Bouwens}, {Illingworth},
  {Labb{\'e}}, {Franx}, {van Dokkum}, {Trenti}, {Stiavelli}, {Gonzalez}, \&
  {Magee}}]{Oesch2013}
{Oesch} P.~A. {et~al.}, 2013, \apj, 773, 75

\bibitem[{{O'Shea} \& {Norman}(2008)}]{Oshea2008}
{O'Shea} B.~W., {Norman} M.~L., 2008, \apj, 673, 14

\bibitem[{{Paardekooper} {et~al}\mbox{.}(2013){Paardekooper}, {Khochfar}, \&
  {Dalla Vecchia}}]{Paardekooper2013}
{Paardekooper} J.-P., {Khochfar} S., {Dalla Vecchia} C., 2013, \mnras, 429, L94

\bibitem[{{Paardekooper} {et~al}\mbox{.}(2011){Paardekooper}, {Pelupessy},
  {Altay}, \& {Kruip}}]{Paardekooper2011}
{Paardekooper} J.-P., {Pelupessy} F.~I., {Altay} G., {Kruip} C.~J.~H., 2011,
  \aap, 530, A87

\bibitem[{{Pawlik} {et~al}\mbox{.}(2009){Pawlik}, {Schaye}, \& {van
  Scherpenzeel}}]{Pawlik2009}
{Pawlik} A.~H., {Schaye} J., {van Scherpenzeel} E., 2009, \mnras, 394, 1812

\bibitem[{{Planck Collaboration}(2014)}]{Planck2013_16}
{Planck Collaboration}, 2014, \aap, 566, A54

\bibitem[{{Planck Collaboration} {et~al}\mbox{.}(2013){Planck Collaboration},
  {Ade}, {Aghanim}, {Armitage-Caplan}, {Arnaud}, {Ashdown}, {Atrio-Barandela},
  {Aumont}, {Baccigalupi}, {Banday}, \& et~al.}]{Planck2013_24}
{Planck Collaboration} {et~al.}, 2013, ArXiv e-prints

\bibitem[{{Robertson} {et~al}\mbox{.}(2013){Robertson}, {Furlanetto},
  {Schneider}, {Charlot}, {Ellis}, {Stark}, {McLure}, {Dunlop}, {Koekemoer},
  {Schenker}, {Ouchi}, {Ono}, {Curtis-Lake}, {Rogers}, {Bowler}, \&
  {Cirasuolo}}]{Robertson2013}
{Robertson} B.~E. {et~al.}, 2013, \apj, 768, 71

\bibitem[{{Ross} {et~al}\mbox{.}(2013){Ross}, {McGreer}, {White}, {Richards},
  {Myers}, {Palanque-Delabrouille}, {Strauss}, {Anderson}, {Shen}, {Brandt},
  {Y{\`e}che}, {Swanson}, {Aubourg}, {Bailey}, {Bizyaev}, {Bovy}, {Brewington},
  {Brinkmann}, {DeGraf}, {Di Matteo}, {Ebelke}, {Fan}, {Ge}, {Malanushenko},
  {Malanushenko}, {Mandelbaum}, {Maraston}, {Muna}, {Oravetz}, {Pan},
  {P{\^a}ris}, {Petitjean}, {Schawinski}, {Schlegel}, {Schneider}, {Silverman},
  {Simmons}, {Snedden}, {Streblyanska}, {Suzuki}, {Weinberg}, \&
  {York}}]{Ross2013}
{Ross} N.~P. {et~al.}, 2013, \apj, 773, 14

\bibitem[{{Schechter}(1976)}]{Schechter1976}
{Schechter} P., 1976, \apj, 203, 297

\bibitem[{{Scoccimarro}(1998)}]{Scoccimarro1998}
{Scoccimarro} R., 1998, \mnras, 299, 1097

\bibitem[{{Shandera} {et~al}\mbox{.}(2011){Shandera}, {Dalal}, \&
  {Huterer}}]{Shandera2011}
{Shandera} S., {Dalal} N., {Huterer} D., 2011, JCAP, 3, 17

\bibitem[{{Silverstein} \& {Tong}(2004)}]{Silverstein2004}
{Silverstein} E., {Tong} D., 2004, \prd, 70, 103505

\bibitem[{{So} {et~al}\mbox{.}(2014){So}, {Norman}, {Reynolds}, \&
  {Wise}}]{So2014}
{So} G.~C., {Norman} M.~L., {Reynolds} D.~R., {Wise} J.~H., 2014, \apj, 789,
  149

\bibitem[{{Springel}(2005)}]{Springel2005}
{Springel} V., 2005, \mnras, 364, 1105

\bibitem[{{Springel} {et~al}\mbox{.}(2001){Springel}, {Yoshida}, \&
  {White}}]{Springel2001}
{Springel} V., {Yoshida} N., {White} S.~D.~M., 2001, \na, 6, 79

\bibitem[{{Tacchella} {et~al}\mbox{.}(2013){Tacchella}, {Trenti}, \&
  {Carollo}}]{Tacchella2013}
{Tacchella} S., {Trenti} M., {Carollo} C.~M., 2013, \apjl, 768, L37

\bibitem[{{Tweed} {et~al}\mbox{.}(2009){Tweed}, {Devriendt}, {Blaizot},
  {Colombi}, \& {Slyz}}]{Tweed2009}
{Tweed} D., {Devriendt} J., {Blaizot} J., {Colombi} S., {Slyz} A., 2009, \aap,
  506, 647

\bibitem[{{Valageas} \& {Nishimichi}(2011)}]{Valageas2011}
{Valageas} P., {Nishimichi} T., 2011, \aap, 527, A87

\bibitem[{{Volonteri} \& {Gnedin}(2009)}]{Volonteri2009}
{Volonteri} M., {Gnedin} N.~Y., 2009, \apj, 703, 2113

\bibitem[{{Wilkins} {et~al}\mbox{.}(2013){Wilkins}, {Bunker}, {Coulton},
  {Croft}, {Matteo}, {Khandai}, \& {Feng}}]{Wilkins2013}
{Wilkins} S.~M., {Bunker} A., {Coulton} W., {Croft} R., {Matteo} T.~D.,
  {Khandai} N., {Feng} Y., 2013, \mnras, 430, 2885

\bibitem[{{Wise} \& {Abel}(2007)}]{Wise2007}
{Wise} J.~H., {Abel} T., 2007, \apj, 671, 1559

\bibitem[{{Wise} \& {Cen}(2009)}]{Wise2009}
{Wise} J.~H., {Cen} R., 2009, \apj, 693, 984

\bibitem[{{Wise} {et~al}\mbox{.}(2014){Wise}, {Demchenko}, {Halicek}, {Norman},
  {Turk}, {Abel}, \& {Smith}}]{Wise2014}
{Wise} J.~H., {Demchenko} V.~G., {Halicek} M.~T., {Norman} M.~L., {Turk} M.~J.,
  {Abel} T., {Smith} B.~D., 2014, ArXiv e-prints

\bibitem[{{Yadav} \& {Wandelt}(2010)}]{Yadav2010}
{Yadav} A.~P.~S., {Wandelt} B.~D., 2010, Advances in Astronomy, 2010

\bibitem[{{Zheng} {et~al}\mbox{.}(2005){Zheng}, {Berlind}, {Weinberg},
  {Benson}, {Baugh}, {Cole}, {Dav{\'e}}, {Frenk}, {Katz}, \&
  {Lacey}}]{Zheng2005}
{Zheng} Z. {et~al.}, 2005, \apj, 633, 791

\end{thebibliography}

%\appendix

\label{lastpage}

%TC:endignore

\end{document}